\pdfoutput=1
\documentclass[aps,prl,reprint,groupedaddress,preprintnumbers]{revtex4-1}
\usepackage[utf8]{inputenc}
\usepackage{amssymb}
\usepackage{amsmath}
\usepackage[colorlinks=true,linkcolor=blue,citecolor=blue, urlcolor=blue]{hyperref}   
\usepackage{tikz}
\usepackage{graphicx}

\definecolor{uibred}{RGB}{167, 38, 47}
\def\Eq#1{Eq.~(\ref{#1})}
\def\Eqs#1{Eqs.~(\ref{#1})}
\def\eq#1{(\ref{#1})}

\def\Fig#1{Fig.~\ref{#1}}

\def\p{\mathbf{p}}

\def\st{\begin{equation}}
\def\stp{\end{equation}}
\def\llangle{\left\langle}
\def\rrangle{\right\rangle}
\usepackage{tikz}

\begin{document}

\title{Chemical Equilibration in Hadronic Collisions}

\author{Aleksi Kurkela}
\email[]{a.k@cern.ch}
\affiliation{Theoretical Physics Department, CERN, Geneva, Switzerland}
\affiliation{Faculty of Science and Technology, University of Stavanger, 
4036 Stavanger, Norway}

\author{Aleksas Mazeliauskas}
\email[]{a.mazeliauskas@thphys.uni-heidelberg.de}
\affiliation{Institut f\"{u}r Theoretische Physik, Universit\"{a}t Heidelberg, 
69120 Heidelberg, Germany}

\date{\today}

\preprint{CERN-TH-2018-238}
\begin{abstract}
We study chemical equilibration in out-of-equilibrium Quark-Gluon Plasma using the first principles method of QCD effective kinetic theory, accurate at weak coupling. 
In longitudinally expanding systems---relevant for relativistic nuclear collisions---we find that for realistic couplings chemical equilibration takes place after hydrodynamization, but well before local thermalization. We estimate that hadronic collisions with final state multiplicities ${dN_\text{ch}}/{d\eta}\gtrsim 10^2$ live long enough to reach approximate chemical equilibrium, which is consistent with the saturation of strangeness enhancement observed in proton-proton, proton-nucleus and  nucleus-nucleus collisions.
\end{abstract} 
\maketitle

The experiments at the LHC and RHIC have seen signs of collective behaviour 
 in proton-proton, proton-nucleus, and nucleus-nucleus collisions, which emerge smoothly as a function of the
system size measured by event multiplicities. 
The signals of collectivity include long range multi-particle correlations~\cite{Abelev:2014mda,Aaboud:2018syf,Aaboud:2017blb,Sirunyan:2017uyl,Aidala:2018mcw,Adare:2018zkb,Adams:2004bi,Adamczyk:2016exq}, indicative of 
the onset of  flow-like phenomena,
and changes in the hadrochemical composition~\cite{Abelev:2007xp,ABELEV:2013zaa,Abelev:2013haa,ALICE:2017jyt}, indicative of modifications to the process of hadronization in dense medium. 
The observed enhancement of (multi-)strange hadron yields with respect to pions seems to be fundamentally at odds with the picture of
hadronic collisions as independent superpositions of individual partons, and which cannot be reproduced by the tuning of the 
standard multipurpose event generators~\cite{Buckley:2011ms}
without the inclusion of significant new elements~\cite{Fischer:2016zzs,Sjostrand:2018xcd}.

In the context of nucleus-nucleus collisions, these observations are understood as signs of kinetically and chemically equilibrated plasma. Fluid dynamic~\cite{Heinz:2013th,Teaney:2009qa,Luzum:2013yya,Gale:2013da,deSouza:2015ena} and statistical hadronization models~\cite{BraunMunzinger:2003zd,Cleymans:2006xj,Andronic:2008gu,Andronic:2017pug} motivated by local thermal and chemical  equilibration of the QGP have enjoyed significant phenomenological success over the past decades in describing  low momentum hadron production  and multiparticle correlations in a range of collision systems. The observed signals of collectivity in small systems, thus, raise the question whether the picture of locally equilibrated plasma can be extended to small systems, and how this picture eventually breaks down. 

How approximately equilibrated plasma emerges from 
fundamental interactions of the medium constituents has been a topic of intense theoretical 
study. There have been significant developments in theoretical understanding of far-from-equilibrium dynamics~\cite{Berges:2013fga,Berges:2013eia}, kinetic equilibration~\cite{Kurkela:2014tea, Kurkela:2015qoa,Keegan:2015avk, Keegan:2016cpi}, and hydrodynamization~\cite{Heller:2015dha,Heller:2016rtz,Strickland:2017kux,Behtash:2017wqg,Romatschke:2017vte} from first principles. These explorations have however limited themselves to gauge theory models that only resemble QCD, such as pure Yang-Mills theory or $\mathcal{N}=4$  supersymmetric Yang-Mills theory, but have not been performed within the full QCD itself%
, where only near-equilibrium dynamics has been studied~\cite{Arnold:2003zc}.
Although for some questions these models can give significant insights, for others---such as chemical equilibration---they lack the essential physics.  The fermion production has been  previously studied using perturbative estimates of collision rates~\cite{Rafelski:1982pu,Shuryak:1992wc}, non-perturbative classical-statistical simulations~\cite{Gelfand:2016prm,Tanji:2017xiw}, solving
rate equations~\cite{Biro:1993qt,Elliott:1999uz}, and
 pQCD based Boltzmann transport models~\cite{Geiger:1991nj,
Borchers:2000wf,Xu:2004mz,Blaizot:2014jna,Ruggieri:2015tsa}.
 In this Letter we address for the first time the emergence of kinetically and chemically equilibrated Quark-Gluon Plasma in full QCD in an \emph{ab initio} framework.  

The setup we employ is the effective QCD kinetic theory of Ref.~\cite{Arnold:2002zm}, with initial conditions set by saturation framework \cite{Lappi:2006fp, Gelis:2010nm, Lappi:2011ju}. This systematically improvable setup is accurate in the asymptotic limit of large center-of-mass energies $\sqrt{s}\rightarrow \infty$. Although the conditions in physical collisions taking place at RHIC and the LHC are probably different  from this idealized limit, the effective theory framework still provides a semi-quantitative physical picture of QCD dynamics that is strongly rooted in the underlying quantum field theory. The kinetic theory encompasses fluid dynamics in the limit of large number of scatterings, but goes beyond the macroscopic fluid dynamical description and can be used to study systems that are far from equilibrium. 

By mapping the only free parameter of the QCD kinetic theory---the coupling constant---to the transport properties in the fluid dynamic limit,  
we see that 
 the system hydrodynamizes quickly in accordance with findings in pure Yang-Mills theory~\cite{Kurkela:2015qoa}. We also find that the chemical equilibration takes place after the system has hydrodynamized but before it finally isotropizes. 
Using the knowledge of the chemical equilibration time, we estimate what is the smallest system that reaches chemical equilibrium and at what multiplicities we expect the hadrochemistry---and therefore also strangeness enhancement---to saturate.

\noindent\textbf{Setup:}  
The kinetic theory that we use to describe equilibration is the 
Effective Kinetic Theory (EKT) of Arnold, Moore, and Yaffe~\cite{Arnold:2002zm}, 
which is leading order accurate in the QCD coupling constant $\lambda= g^2 N_c=  4\pi \alpha_s N_c$.
This framework includes  Hard Thermal Loop (HTL)  in-medium screening effects~\cite{Braaten:1989mz}, and Landau-Pomeranchuk-Migdal~\cite{Landau:1953gr, Landau:1953um,Migdal:1956tc,Migdal:1955nv} suppression of
collinear radiation in
far-from-equilibrium, but parametrically isotropic systems~\footnote{Anisotropic systems suffer from the presence of unstable plasma modes~\cite{Mrowczynski:1988dz, Mrowczynski:1993qm,Mrowczynski:2000ed}, which could change the kinetic dynamics~\cite{Kurkela:2011ub,Kurkela:2011ti}. However detailed 3+1D classical-statistical YM simulations found no effects of plasma instabilities beyond very early times~\cite{Berges:2013fga,Berges:2013eia}. Therefore we will use isotropic approximations, which remove the unstable modes from the kinetic description. Note that there are no unstable fermionic modes~\cite{Mrowczynski:2001az,Schenke:2006fz}}.  We extend the previously developed setup of Refs.~\cite{Kurkela:2014tea, Kurkela:2015qoa,Keegan:2015avk, Keegan:2016cpi} by including quark degrees of freedom (see our  companion paper for more details~\cite{Kurkela:2018oqw}).

We numerically solve the Boltzmann equation for homogeneous boost invariant quark and gluon distribution functions $f_{g,q}$ according to
 \begin{align}
\partial_\tau f_s(\p,\tau) - \frac{p^z}{\tau }\partial_{p^z}f_s(\p, \tau)
 &= -\mathcal{C}^s_{2\leftrightarrow 2}[f]-\mathcal{C}^s_{1\leftrightarrow 2}[f],\label{eq:bolz}
 \end{align}
where $\tau$ is the Bjorken time $\tau=\sqrt{t^2-z^2}$~\cite{Bjorken:1982qr}.
The collision kernel $C_{2\rightarrow2}^s[f](\p,\tau)$ is a multidimensional integral over the $2\leftrightarrow 2$ scattering matrix elements $|\mathcal{M}^{ab}_{cd}|^2$ and phase-space factors, which
 describes the scattering rates for $gg\leftrightarrow gg$, $gq\leftrightarrow gq$, $qq\leftrightarrow qq$ and  $gg\leftrightarrow q\bar q$ processes~\cite{Arnold:2002zm}.  
For soft small angle scatterings the tree level scattering matrix $|\mathcal{M}^{ab}_{cd}|^2$ is divergent and in-medium screening effects must be computed using the HTL resumed propagators.
In practice, we supplement the divergent terms appearing in soft gluon 
or fermion exchanges, \emph{e.g.} ${(u-s)}/{t}\sim {1}/{q^2}$, with an infrared regulator~\cite{York:2014wja}
\begin{equation}
\frac{u-s}{t} \rightarrow\frac{u-s}{t}\frac{q^2}{q^2+\xi_s^2 m_s^2}\label{eq:reg},
\end{equation}
where  $q = |\p' - \p|$ is the momentum transfer in $t$-channel, and
where $m_{g,q}$ are the in-medium screening masses~\cite{Arnold:2002zm}.
Constants $\xi_g = e^{5/6}/2 $ and $\xi_q=e/2$ are fixed such 
that the matrix elements 
reproduce the full HTL results in isotropic systems for
drag and momentum diffusion properties of soft gluon scattering~\cite{York:2014wja}
and gluon to quark conversion $gg\rightarrow q\bar q$~\cite{Ghiglieri:2015ala,teaney}.

The particle number changing processes $g\leftrightarrow gg$, $q\leftrightarrow qg$, $g\leftrightarrow q\bar q$ are included in $\mathcal{C}^s_{1\leftrightarrow 2}[f]( \p, \tau)$ collision kernel~\cite{Arnold:2002zm}.
 The effective splitting rates are calculated using  an isotropic screening approximation~\cite{Aurenche:2002pd}.

We use
 Color-Glass-Condensate motivated
  initial conditions for the gluon distribution function~\cite{Lappi:2006fp, Gelis:2010nm}, which have also been studied in pure gauge theory in \cite{Kurkela:2015qoa, Keegan:2016cpi,Kurkela:2018vqr,Kurkela:2018wud}. Specifically at $\tau_0=1/Q_s$ we take
\begin{align}
f^g({\bf p}, \tau= \tau_0)=\frac{2 A}{\lambda} \frac{Q_0}{\sqrt{p_\perp^2+p_z^2\xi^2}} e^{-\frac{2}{3} \frac{p_\perp^2+\xi^2 p_z^2}{Q_0^2}},\label{eq:focc}
\end{align}
where the values of $A$ and $Q_0$, and $\xi$ are adjusted to reproduce the typical  transverse momentum and energy density  from the classical lattice simulations of initial stages of the collision~\cite{Kurkela:2015qoa,Lappi:2011ju}. 
Within the saturation framework, the energy density of fermions is parametrically suppressed compared to that of the gluons and consistently we set the initial fermion energy density to zero in the following. In addition, we have checked that starting with small but non-zero fermion energy density does not change the conclusions.

In collisions with realistic center of mass energies, the QCD coupling constant for in-medium energy scale is not small  $\alpha_s(Q_s\sim 1\,\text{GeV})\gtrsim 0.3$~\cite{Tanabashi:2018oca}. At such values one expects higher order corrections to the macroscopic medium properties. Indeed, NLO calculations of specific shear viscosity $\eta/s$ show sizeable modifications of leading order results~\cite{Ghiglieri:2018dib}; however, other transport coefficients in units of $\eta/s$, e.g. $\tau_\pi/(\eta/(sT))$, are less sensitive to the changes of the coupling constant~\cite{Ghiglieri:2018dgf}.  Therefore by adjusting the coupling constant $\lambda$ to reproduce given specific viscosity $\eta/s(\lambda)$~\footnote{We determine $\eta/s(\lambda)$ numerically from near-equilibrium pressure anisotropy evolution in longitudinally expanding system~\cite{Kurkela:2018oqw}.} we fix the only unspecified parameter of our model and set the overall speed of microscopic dynamics. In practice, we perform simulations for multiple values of $\lambda$ and map our results to physical values of $\eta/s$ extracted from other models. 

\begin{figure}
\includegraphics[width=0.9\linewidth]{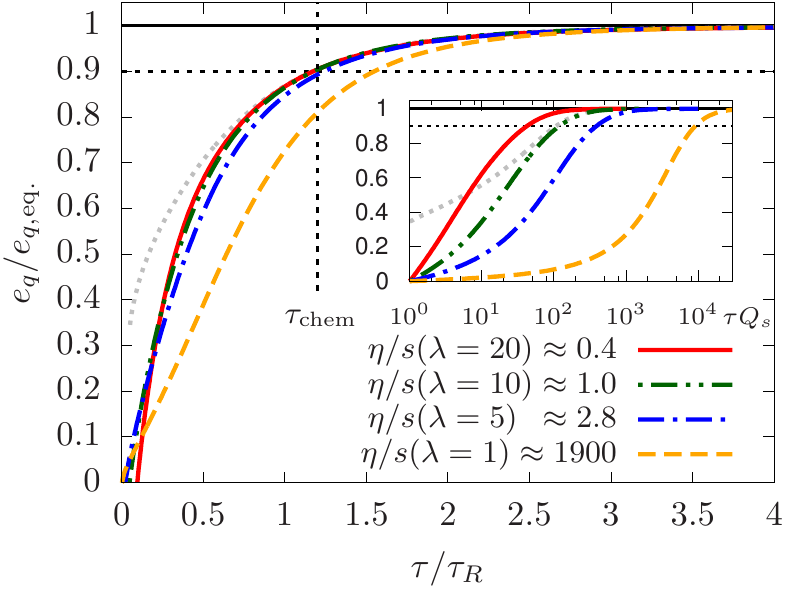}
\caption{Fermion energy density  fraction of equilibrium density $e_q(\tau)/e_{q,\text{eq}}(\tau)$ as a function of rescaled time $\tau/\tau_R=\tau T_\text{id.}/(4\pi\eta/s)$ for different coupling constants $\lambda=1,5,10,20$. The inset shows un-rescaled time dependence on a log-time plot.\label{fig:eq_var} Grey dotted line shows evolution with non-zero initial fermion density.}
\end{figure}

\noindent\textbf{Results\label{sec:exp}:}
Starting from 
initial conditions of \Eq{eq:focc}, we solve the $N_c=3$ QCD transport equation \Eq{eq:bolz} for $N_f=3$ flavours of massless fermions. 
The dynamics of a near-equilibrium system at temperature $T$ is governed by the kinetic relaxation time, or mean free path,
\begin{equation}
\tau_R = \frac{4\pi \eta}{s T} \sim \frac{1}{\lambda^2 T}\label{eq:tauR1}.
\end{equation}
At late times when the system is close to local thermal equilibrium, the time evolution of the temperature is given by ideal hydrodynamics with constant $T(\tau)\tau^{1/3}$. 
We follow the practice of \cite{Kurkela:2018vqr,Kurkela:2018wud} and
  find what the temperature of the system would have been at earlier times if the full evolution of the system was described by ideal fluid dynamics
\begin{align}
T_{\rm id.}(\tau) = \frac{(T(\tau)\tau^{1/3})|_{\tau \rightarrow \infty}}{\tau^{1/3}},\label{eq:Tid}
\end{align}
and define a time dependent kinetic relaxation time
$\tau_R(\tau) = \frac{4\pi \eta/s}{T_{\rm id.}(\tau)}\label{eq:tauRexp},$
which we use to compare simulations with different $\eta/s$.

The total energy density in a multi-component plasma is given by a sum of its parts
\begin{align}
e(\tau) &= \int\! \frac{d^3 p}{(2\pi)^3} p^0 (\nu_g f_g+ 2 N_f\nu_q f_q) = e_g + e_q,
\end{align}
where $\nu_g = 2(N_c^2-1)$ and $\nu_q=2N_c$. In chemical equilibrium  the fermion content of the plasma constitutes $r_q\approx 0.66$ fraction of the total density, i.e. $e_{q, \text{eq}} \equiv r_q e$, where $r_q^{-1}=1+\frac{\nu_g}{2N_f\nu_q}\frac{8}{7}$.
We study how the equilibrium fermion energy fraction is reached in \Fig{fig:eq_var} for different values of the coupling constant $\lambda$. 
We see that expressing time in units of kinetic relaxation time $\tau/\tau_R$ reduces the vast separation of equilibration timescales as shown in the inset plot. For coupling constants $\lambda=5,10,20$ corresponding to $\alpha_s\approx 0.1-0.5$, the chemical equilibration becomes approximately universal. At these moderate couplings, the 90\% of fermion equilibrium energy fraction is reached at time $\tau\approx 1.2\tau_R$, which 
we take as our somewhat arbitrary definition of the chemical equilibration time, \emph{i.e.},
\begin{equation} 
\left|1-\frac{e_q(\tau_\text{chem})}{e_{q,{\rm eq}}}\right| = 0.1, \textrm{ with }\quad \tau_\text{chem} = 1.2 \tau_R.\label{eq:tauchem}
\end{equation}

\begin{figure}
\centering
{\includegraphics[width=0.9\linewidth]{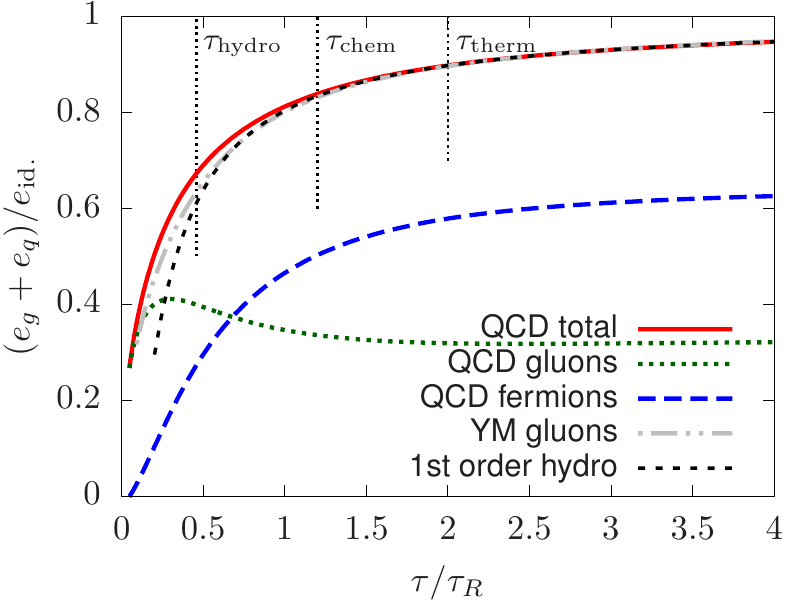}}
\caption{The total energy density evolution in QCD   kinetic theory (red solid line)  scaled by ideal asymptotics $e_\text{id.}=\frac{\pi^2}{30}(\nu_g+\frac{7}{8}\nu_g)T_\text{id.}^4$ for $(\eta/s)_\text{QCD}(\lambda=10) \approx 1.0$.
The gluonic and fermionic energy components are shown by green dotted and blue dashed lines correspondingly. In addition energy evolution in Yang-Mills kinetic theory for the same initial conditions is shown by grey dash-dotted line ($(\eta/s)_\text{YM}\approx0.62$). 
}
\label{hydro}
\end{figure}

To quantify the approach to thermal equilibrium and hydrodynamization we define two additional timescales $\tau_\text{therm}$ and $\tau_\text{hydro}$, similarly to \Eq{eq:tauchem}, by requiring the combined gluon and fermion energy density $e=e_g+e_q$ to be within 10\% of ideal and viscous hydrodynamic estimates of energy density
\begin{align} 
\left|1-\frac{e(\tau_\text{therm})}{e_{{\rm id.}}}\right| = 0.1,\quad
\left|1-\frac{e(\tau_\text{hydro})}{e_{{\rm 1st}}}\right| = 0.1 \label{eq:hydro}\,.
\end{align}
Here $e_{{\rm id.}} =\frac{\pi^2}{30} (\nu_g + \frac{7}{8}\nu_q 2N_f) T^4_\text{id}$ is the ideal estimate of the energy density, and 
$e_{{\rm 1st}} =\frac{\pi^2}{30} (\nu_g + \frac{7}{8}\nu_q 2N_f) T^4_\text{1st}$, where
 $T_\text{1st}$ is the 1st order viscous hydrodynamic solution of longitudinally expanding system
~\cite{Kouno:1989ps,Muronga:2001zk}
\begin{equation}
{T_\text{1st}(\tau)}=T_\text{id}(\tau)\left(1-\frac{2}{12\pi}\frac{\tau_R}{\tau}\right).\label{eq:T1st}
\end{equation}

In \Fig{hydro} we show the evolution of the total energy density for full QCD kinetic theory (solid red line) for $\lambda=10$ ($\alpha_s\approx0.26$) as a fraction of ideal energy density $e_\text{id.}(\tau)$. We see that the system rapidly approaches hydrodynamic behaviour and at $\tau_\text{hydro}\approx 0.46\tau_R$ the total energy density is within $10\%$ of energy density given by viscous hydrodynamic evolution (black dotted line)~\footnote{Note that the definition of $\tau_\text{hydro}$ in \Eq{eq:hydro} differs from the one used in Ref.~\cite{Kurkela:2018vqr,Kurkela:2018wud}.}. The ideal limit is approached only very slowly and thermalization takes place at much later times $\tau_\text{therm}\approx 2$. 
In the meantime the chemical composition of plasma undergoes a rapid conversion and  the energy density stored in quark degrees  of freedom (blue dashed line) quickly overtakes the gluonic component  (green dotted line). 
This results in the following ordering of equilibration time-scales
\begin{equation}
\tau_\text{hydro}<\tau_\text{chem}< \tau_\text{therm},
\end{equation}
according to  the criteria given by \Eqs{eq:tauchem} and \eq{eq:hydro}.

We also compare the total energy density evolution in QCD and pure Yang-Mills kinetic theory (grey dashed line) in \Fig{hydro}. 
After rescaling with corresponding kinetic relaxation time $\tau_R$ and temperature $T_\text{id.}$, the total energy density evolution is rather similar in both Yang-Mills and QCD kinetic theories. This justifies \emph{a posteriori} the use of pure gauge theory in modelling of the hydrodynamization in nuclear collisions in \cite{Kurkela:2018vqr,Kurkela:2018wud}. 
Finally, we comment in passing that starting with small, but non-zero initial fermion density, does not change the chemical  equilibration time as demonstrated by a grey dotted line in \Fig{fig:eq_var} for initial fermion to gluon energy fraction $e_q/e_{g}=0.3$.

\noindent\textbf{Discussion\label{sec:mult}:}
As seen in \Fig{fig:eq_var}, the process of chemical equilibration becomes insensitive to the value of the coupling constant when measured in properly scaled units. We may use this insensitivity to extrapolate our results to conditions expected to take place in physical collisions at hadron colliders.  By taking realistic values of $\left.(\tau^{1/3} T)\right|_\infty$ and $\eta/s$ estimated from hydrodynamical analysis, we convert dimensionless time $\tau/\tau_R$  into $\text{fm}/c$.
This gives us a unique prediction  based on first principle QCD kinetic theory for the early time evolution where the fluid dynamical description is not valid. It is a non-trivial question whether such pre-equilibrium evolution 
will be consistent with the subsequent fluid dynamical evolution of thermally and chemically equilibrated QGP.

Following the procedure presented in \cite{Kurkela:2018wud}, the asymptotic constant $\left.(\tau^{1/3} T)\right|_\infty$ in kinetic theory can be fixed by  the averaged entropy density per rapidity in hydrodynamic simulations
\begin{equation}
\left.(\tau T^3)\right|_\infty = {\left<s\tau\right>}/ {\left(\tfrac{4\pi^2}{90}\nu_\text{eff}\right)},
\end{equation}
which is a robust quantity and agrees well between different hydrodynamic implementations~\cite{Keegan:2016cpi,Shen:2014vra,Niemi:2015qia}. 
Then $\tau/\tau_R=\tau T_\text{id}/(4 \pi \eta/s)$ can be inverted to express physical time in terms of scaled time and asymptotic constants
\begin{equation}
\tau= \left({\tau}/{\tau_R}\right)^{3/2}(4\pi\eta/s)^{3/2} \left<s\tau\right>^{-1/2} \left(\tfrac{4\pi^2}{90}\nu_\text{eff}\right)^{1/2}.\label{eq:conv}
\end{equation}
We  take  $\llangle \tau s\rrangle\approx 4.1 \, {\rm  GeV}^2 $ as a typical value from hydrodynamic simulations for central Pb-Pb collisions at $\sqrt{s_{NN}}=2.72\,\text{TeV}$~\cite{Keegan:2016cpi} and $\nu_{\rm
eff}(0.4\,\text{GeV})\approx 40$ as the effective number of degrees of freedom obtained from lattice QCD~\cite{Bazavov:2014pvz,Borsanyi:2016ksw} (for ideal gas of quarks and gluons $\nu_\text{eff}=47.5$).
The specific shear viscosity extracted from comparison of hydrodynamic models and experimental data vary roughly in a range $\eta/s\approx 0.1{-}0.2$~\cite{Bernhard:2016tnd,Niemi:2015qia} and we take $\eta/s=2/(4\pi)\approx0.16$, which was also used in \cite{Kurkela:2018wud}. 

\begin{figure}
\includegraphics[width=0.9\linewidth]{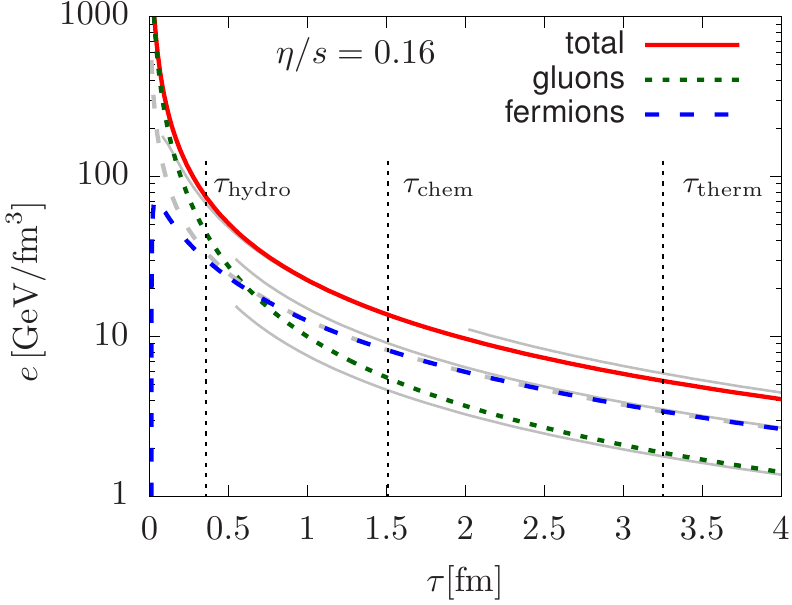}
\caption{Evolution of total energy density and its gluonic and fermion components in kinetic theory converted to physical units using universality of $\tau/\tau_R$ scaling and physical values of $\eta/s=0.16$, $\left<s\tau\right>=4.1\,\text{GeV}^2$ and $\nu_\text{eff}=40$. The grey solid lines 
correspond to ideal, viscous and chemically equilibrated energies. Grey dashed line corresponds to fermion energy evolution with non-zero initial fermion density.
The time axis dependence on asymptotic constants is given by $\tau [\text{fm}]\times (\eta/s/0.16)^{3/2}({\left<s\tau\right>}/{4.1\,\text{GeV}^2})^{-1/2}({\nu_\text{eff}}/{40})^{1/2}$, whereas the energy axis scales as
$e[\text{GeV/fm}^3]\times (\eta/s/0.16)^{-2}({\left<s\tau\right>}/{4.1\,\text{GeV}^2})^{2}({\nu_\text{eff}}/{40})^{-1}$. \label{fig:conv}}
\end{figure}

In \Fig{fig:conv} we show the energy evolution of \Fig{hydro} now converted to physical units according to \Eq{eq:conv}. 
We see that starting with initial conditions with no fermions, the fermion energy density increases rapidly to its maximum value at very early times. When this maximum is reached depends sensitively on the initial conditions of the fermions, and for non-zero initial fermion distribution the energy density can even decrease monotonically (gray dashed line). 
Although this uncertainty does not affect the equilibration times, there are observables, such as photon or dilepton production, that may be sensitive to the initial fermion fraction at early times.
At later times, the fermionic energy density decreases slower than the gluonic component thus increasing the fermion fraction. At times $\tau\sim 0.7\,\text{fm}$ fermion energy density starts to dominate and  $\tau \sim 1.5\,\text{fm}$ fermionic energy fraction is within $10\%$ of equilibrium value shown by grey line. 
Finally, if the system continues expanding with boost and transverse translation invariance then at $\tau\sim3.3\,\text{fm}$ energy density evolution is within 10\% of the ideal hydrodynamic expectation and the system can be consider locally thermalized.

However, the approximation of transverse translational invariance breaks when the central parts of the collision come into causal contact with the edge of the fireball and the system starts to undergo significant radial expansion. 
Following the logic of Ref.~\cite{Kurkela:2018wud}, we assume that the system disintegrates once its lifetime exceeds $\tau\sim R$ and a three dimensional expansion begins, where $R$ is the initial transverse radius of the system~\footnote{For more accurate description, extensions of our study including transverse expansion are needed~\cite{Kurkela:2018qeb}.}.
The charged particle multiplicity $dN_\text{\rm ch}/d\eta$ in the final state can be related to the entropy density according to
 $\langle\tau s\rangle \approx (S/N_\text{\rm ch})~1/A_\bot~dN_\text{\rm ch}/d\eta$, where $A_{\bot}\approx \pi R^2$
and 
$S/N_\text{\rm ch}\approx 7$ is a constant of hadron gas~\cite{Muller:2005en,Hanusthesis}.
 We can now ask what is the minimal system multiplicity that can achieve chemical equilibration before freezing-out.
Rewriting \Eq{eq:conv} as a bound on multiplicities, we get
\begin{equation}
 \frac{dN_{\rm ch}}{d\eta}  \gtrsim 110 \left(\frac{\tau_\text{chem}}{1.2\tau_R}\right)^3 \left(\frac{ \eta/s}{0.16}\right)^3\left(\frac{\tau_\text{chem}}{R}\right)^{-2}\label{eq:nch},
\end{equation}
where other constants were set to their nominal values~\footnote{Note that the combination $ \left({\tau}/{\tau_R}\right)^{3}  \left({\tau}/{R}\right)^{-2} $ is independent of time.}. That is,  using the equilibration rates of QCD kinetic theory, we estimate that chemically equilibrated QGP with specific shear-viscosity $\eta/s= 0.16$ can be formed only for systems with multiplicity  $ {dN_{\rm ch}}/{d\eta}  \gtrsim 10^2$ by the time it starts to freeze-out at $\tau \sim R$. 

Experimental measurements of strangeness enhancement in p-p,   p-Pb, and Pb-Pb collisions clearly indicate a continuous increase of strangeness production, which is saturated around ${dN_{\rm ch}}/{d\eta} \sim 10^2$~\cite{ALICE:2017jyt,Abelev:2013haa}.
Our calculation does not contain the necessary ingredients to describe the hadrochemistry in detail~\footnote{For example, our model  neglects quark masses, which could delay flavour equilibration. Even then the relation of strangeness content in the QGP and hadronic phase is not trivial~\cite{Lee:1987mj,Sollfrank:1995bn,Letessier:1993hi}}. However,  assuming that the underlying physics that saturates the strangeness production is the formation of chemically equilibrated plasma, our chemical equilibration rate gives a necessary condition for the event multiplicities where 
the saturation can take place. 
We 
note that as \Eq{eq:nch} does not depend on the physical size of the system other than in the combination $\tau_{\rm chem}/R\sim 1$, 
 our calculation is 
 consistent with the observed overlap of strangeness enhancement across different collision systems, when plotted as a function of multiplicity.
For $\eta/s =0.16$, our estimate of the multiplicity where we expect strangeness saturation to take place roughly agrees with the experimental observation.
 Although our estimate is strongly dependent on the definition of $\tau_\text{chem}$ and the assumption that the process of equilibration terminates at $\tau_\text{chem}/R\sim 1$, it still seems that large values of $\eta/s\sim 1$ would  be in contradiction with chemically equilibrated QGP for  ${dN_{\rm ch}}/{d\eta} \sim 10^2$ thanks to the strong dependence on specific shear viscosity in  \Eq{eq:nch}.
In conclusion, in this novel way, we connect the dynamical transport properties of Quark-Gluon Plasma to hadrochemical output. Equation
\eq{eq:nch} predicts that the saturation of strangeness enhancement should be observed at the same final state multiplicity across different collision systems. 
We expect that the future high-luminosity studies of proton-proton, proton-nucleus, and nucleus-nucleus collisions at the LHC will answer this question conclusively \cite{Citron:2018lsq}.

\noindent \textbf{Acknowledgements:} 
The authors thank  Peter Arnold, J\"urgen Berges, Ulrich Heinz, Jacopo Ghiglieri, Jean-Fran\c cois Paquet, S\"oren Schlichting, Derek Teaney, and Urs Wiedemann for valuable discussions.
This work was supported in part by the German Research Foundation (DFG) 
Collaborative Research Centre (SFB) 1225 (ISOQUANT) (A.M.). Finally, A.M.
 thanks CERN Theoretical Physics
Department for the hospitality during the short-term visit.

\bibliography{master.bib}

\begin{thebibliography}{94}%
\makeatletter
\providecommand \@ifxundefined [1]{%
 \@ifx{#1\undefined}
}%
\providecommand \@ifnum [1]{%
 \ifnum #1\expandafter \@firstoftwo
 \else \expandafter \@secondoftwo
 \fi
}%
\providecommand \@ifx [1]{%
 \ifx #1\expandafter \@firstoftwo
 \else \expandafter \@secondoftwo
 \fi
}%
\providecommand \natexlab [1]{#1}%
\providecommand \enquote  [1]{``#1''}%
\providecommand \bibnamefont  [1]{#1}%
\providecommand \bibfnamefont [1]{#1}%
\providecommand \citenamefont [1]{#1}%
\providecommand \href@noop [0]{\@secondoftwo}%
\providecommand \href [0]{\begingroup \@sanitize@url \@href}%
\providecommand \@href[1]{\@@startlink{#1}\@@href}%
\providecommand \@@href[1]{\endgroup#1\@@endlink}%
\providecommand \@sanitize@url [0]{\catcode `\\12\catcode `\$12\catcode
  `\&12\catcode `\#12\catcode `\^12\catcode `\_12\catcode `\%12\relax}%
\providecommand \@@startlink[1]{}%
\providecommand \@@endlink[0]{}%
\providecommand \url  [0]{\begingroup\@sanitize@url \@url }%
\providecommand \@url [1]{\endgroup\@href {#1}{\urlprefix }}%
\providecommand \urlprefix  [0]{URL }%
\providecommand \Eprint [0]{\href }%
\providecommand \doibase [0]{http://dx.doi.org/}%
\providecommand \selectlanguage [0]{\@gobble}%
\providecommand \bibinfo  [0]{\@secondoftwo}%
\providecommand \bibfield  [0]{\@secondoftwo}%
\providecommand \translation [1]{[#1]}%
\providecommand \BibitemOpen [0]{}%
\providecommand \bibitemStop [0]{}%
\providecommand \bibitemNoStop [0]{.\EOS\space}%
\providecommand \EOS [0]{\spacefactor3000\relax}%
\providecommand \BibitemShut  [1]{\csname bibitem#1\endcsname}%
\let\auto@bib@innerbib\@empty
\bibitem [{\citenamefont {Abelev}\ \emph
  {et~al.}(2014{\natexlab{a}})\citenamefont {Abelev} \emph
  {et~al.}}]{Abelev:2014mda}%
  \BibitemOpen
  \bibfield  {author} {\bibinfo {author} {\bibfnamefont {B.~B.}\ \bibnamefont
  {Abelev}} \emph {et~al.} (\bibinfo {collaboration} {ALICE}),\ }\href
  {\doibase 10.1103/PhysRevC.90.054901} {\bibfield  {journal} {\bibinfo
  {journal} {Phys. Rev.}\ }\textbf {\bibinfo {volume} {C90}},\ \bibinfo {pages}
  {054901} (\bibinfo {year} {2014}{\natexlab{a}})},\ \Eprint
  {http://arxiv.org/abs/1406.2474} {arXiv:1406.2474 [nucl-ex]} \BibitemShut
  {NoStop}%
\bibitem [{\citenamefont {Aaboud}\ \emph {et~al.}(2019)\citenamefont {Aaboud}
  \emph {et~al.}}]{Aaboud:2018syf}%
  \BibitemOpen
  \bibfield  {author} {\bibinfo {author} {\bibfnamefont {M.}~\bibnamefont
  {Aaboud}} \emph {et~al.} (\bibinfo {collaboration} {ATLAS}),\ }\href
  {\doibase 10.1016/j.physletb.2018.11.065} {\bibfield  {journal} {\bibinfo
  {journal} {Phys. Lett.}\ }\textbf {\bibinfo {volume} {B789}},\ \bibinfo
  {pages} {444} (\bibinfo {year} {2019})},\ \Eprint
  {http://arxiv.org/abs/1807.02012} {arXiv:1807.02012 [nucl-ex]} \BibitemShut
  {NoStop}%
\bibitem [{\citenamefont {Aaboud}\ \emph {et~al.}(2018)\citenamefont {Aaboud}
  \emph {et~al.}}]{Aaboud:2017blb}%
  \BibitemOpen
  \bibfield  {author} {\bibinfo {author} {\bibfnamefont {M.}~\bibnamefont
  {Aaboud}} \emph {et~al.} (\bibinfo {collaboration} {ATLAS}),\ }\href
  {\doibase 10.1103/PhysRevC.97.024904} {\bibfield  {journal} {\bibinfo
  {journal} {Phys. Rev.}\ }\textbf {\bibinfo {volume} {C97}},\ \bibinfo {pages}
  {024904} (\bibinfo {year} {2018})},\ \Eprint
  {http://arxiv.org/abs/1708.03559} {arXiv:1708.03559 [hep-ex]} \BibitemShut
  {NoStop}%
\bibitem [{\citenamefont {Sirunyan}\ \emph {et~al.}(2018)\citenamefont
  {Sirunyan} \emph {et~al.}}]{Sirunyan:2017uyl}%
  \BibitemOpen
  \bibfield  {author} {\bibinfo {author} {\bibfnamefont {A.~M.}\ \bibnamefont
  {Sirunyan}} \emph {et~al.} (\bibinfo {collaboration} {CMS}),\ }\href
  {\doibase 10.1103/PhysRevLett.120.092301} {\bibfield  {journal} {\bibinfo
  {journal} {Phys. Rev. Lett.}\ }\textbf {\bibinfo {volume} {120}},\ \bibinfo
  {pages} {092301} (\bibinfo {year} {2018})},\ \Eprint
  {http://arxiv.org/abs/1709.09189} {arXiv:1709.09189 [nucl-ex]} \BibitemShut
  {NoStop}%
\bibitem [{\citenamefont {Aidala}\ \emph {et~al.}(2019)\citenamefont {Aidala}
  \emph {et~al.}}]{Aidala:2018mcw}%
  \BibitemOpen
  \bibfield  {author} {\bibinfo {author} {\bibfnamefont {C.}~\bibnamefont
  {Aidala}} \emph {et~al.} (\bibinfo {collaboration} {PHENIX}),\ }\href
  {\doibase 10.1038/s41567-018-0360-0} {\bibfield  {journal} {\bibinfo
  {journal} {Nature Phys.}\ }\textbf {\bibinfo {volume} {15}},\ \bibinfo
  {pages} {214} (\bibinfo {year} {2019})},\ \Eprint
  {http://arxiv.org/abs/1805.02973} {arXiv:1805.02973 [nucl-ex]} \BibitemShut
  {NoStop}%
\bibitem [{\citenamefont {Adare}\ \emph {et~al.}(2019)\citenamefont {Adare}
  \emph {et~al.}}]{Adare:2018zkb}%
  \BibitemOpen
  \bibfield  {author} {\bibinfo {author} {\bibfnamefont {A.}~\bibnamefont
  {Adare}} \emph {et~al.} (\bibinfo {collaboration} {PHENIX}),\ }\href
  {\doibase 10.1103/PhysRevC.99.024903} {\bibfield  {journal} {\bibinfo
  {journal} {Phys. Rev.}\ }\textbf {\bibinfo {volume} {C99}},\ \bibinfo {pages}
  {024903} (\bibinfo {year} {2019})},\ \Eprint
  {http://arxiv.org/abs/1804.10024} {arXiv:1804.10024 [nucl-ex]} \BibitemShut
  {NoStop}%
\bibitem [{\citenamefont {Adams}\ \emph {et~al.}(2005)\citenamefont {Adams}
  \emph {et~al.}}]{Adams:2004bi}%
  \BibitemOpen
  \bibfield  {author} {\bibinfo {author} {\bibfnamefont {J.}~\bibnamefont
  {Adams}} \emph {et~al.} (\bibinfo {collaboration} {STAR}),\ }\href {\doibase
  10.1103/PhysRevC.72.014904} {\bibfield  {journal} {\bibinfo  {journal} {Phys.
  Rev.}\ }\textbf {\bibinfo {volume} {C72}},\ \bibinfo {pages} {014904}
  (\bibinfo {year} {2005})},\ \Eprint {http://arxiv.org/abs/nucl-ex/0409033}
  {arXiv:nucl-ex/0409033 [nucl-ex]} \BibitemShut {NoStop}%
\bibitem [{\citenamefont {Adamczyk}\ \emph {et~al.}(2016)\citenamefont
  {Adamczyk} \emph {et~al.}}]{Adamczyk:2016exq}%
  \BibitemOpen
  \bibfield  {author} {\bibinfo {author} {\bibfnamefont {L.}~\bibnamefont
  {Adamczyk}} \emph {et~al.} (\bibinfo {collaboration} {STAR}),\ }\href
  {\doibase 10.1103/PhysRevLett.116.112302} {\bibfield  {journal} {\bibinfo
  {journal} {Phys. Rev. Lett.}\ }\textbf {\bibinfo {volume} {116}},\ \bibinfo
  {pages} {112302} (\bibinfo {year} {2016})},\ \Eprint
  {http://arxiv.org/abs/1601.01999} {arXiv:1601.01999 [nucl-ex]} \BibitemShut
  {NoStop}%
\bibitem [{\citenamefont {Abelev}\ \emph {et~al.}(2008)\citenamefont {Abelev}
  \emph {et~al.}}]{Abelev:2007xp}%
  \BibitemOpen
  \bibfield  {author} {\bibinfo {author} {\bibfnamefont {B.~I.}\ \bibnamefont
  {Abelev}} \emph {et~al.} (\bibinfo {collaboration} {STAR}),\ }\href {\doibase
  10.1103/PhysRevC.77.044908} {\bibfield  {journal} {\bibinfo  {journal} {Phys.
  Rev.}\ }\textbf {\bibinfo {volume} {C77}},\ \bibinfo {pages} {044908}
  (\bibinfo {year} {2008})},\ \Eprint {http://arxiv.org/abs/0705.2511}
  {arXiv:0705.2511 [nucl-ex]} \BibitemShut {NoStop}%
\bibitem [{\citenamefont {Abelev}\ \emph
  {et~al.}(2014{\natexlab{b}})\citenamefont {Abelev} \emph
  {et~al.}}]{ABELEV:2013zaa}%
  \BibitemOpen
  \bibfield  {author} {\bibinfo {author} {\bibfnamefont {B.~B.}\ \bibnamefont
  {Abelev}} \emph {et~al.} (\bibinfo {collaboration} {ALICE}),\ }\href
  {\doibase 10.1016/j.physletb.2014.05.052, 10.1016/j.physletb.2013.11.048}
  {\bibfield  {journal} {\bibinfo  {journal} {Phys. Lett.}\ }\textbf {\bibinfo
  {volume} {B728}},\ \bibinfo {pages} {216} (\bibinfo {year}
  {2014}{\natexlab{b}})},\ \bibinfo {note} {[Erratum: Phys.
  Lett.B734,409(2014)]},\ \Eprint {http://arxiv.org/abs/1307.5543}
  {arXiv:1307.5543 [nucl-ex]} \BibitemShut {NoStop}%
\bibitem [{\citenamefont {Abelev}\ \emph
  {et~al.}(2014{\natexlab{c}})\citenamefont {Abelev} \emph
  {et~al.}}]{Abelev:2013haa}%
  \BibitemOpen
  \bibfield  {author} {\bibinfo {author} {\bibfnamefont {B.~B.}\ \bibnamefont
  {Abelev}} \emph {et~al.} (\bibinfo {collaboration} {ALICE}),\ }\href
  {\doibase 10.1016/j.physletb.2013.11.020} {\bibfield  {journal} {\bibinfo
  {journal} {Phys. Lett.}\ }\textbf {\bibinfo {volume} {B728}},\ \bibinfo
  {pages} {25} (\bibinfo {year} {2014}{\natexlab{c}})},\ \Eprint
  {http://arxiv.org/abs/1307.6796} {arXiv:1307.6796 [nucl-ex]} \BibitemShut
  {NoStop}%
\bibitem [{\citenamefont {Adam}\ \emph {et~al.}(2017)\citenamefont {Adam} \emph
  {et~al.}}]{ALICE:2017jyt}%
  \BibitemOpen
  \bibfield  {author} {\bibinfo {author} {\bibfnamefont {J.}~\bibnamefont
  {Adam}} \emph {et~al.} (\bibinfo {collaboration} {ALICE}),\ }\href {\doibase
  10.1038/nphys4111} {\bibfield  {journal} {\bibinfo  {journal} {Nature Phys.}\
  }\textbf {\bibinfo {volume} {13}},\ \bibinfo {pages} {535} (\bibinfo {year}
  {2017})},\ \Eprint {http://arxiv.org/abs/1606.07424} {arXiv:1606.07424
  [nucl-ex]} \BibitemShut {NoStop}%
\bibitem [{\citenamefont {Buckley}\ \emph {et~al.}(2011)\citenamefont {Buckley}
  \emph {et~al.}}]{Buckley:2011ms}%
  \BibitemOpen
  \bibfield  {author} {\bibinfo {author} {\bibfnamefont {A.}~\bibnamefont
  {Buckley}} \emph {et~al.},\ }\href {\doibase 10.1016/j.physrep.2011.03.005}
  {\bibfield  {journal} {\bibinfo  {journal} {Phys. Rept.}\ }\textbf {\bibinfo
  {volume} {504}},\ \bibinfo {pages} {145} (\bibinfo {year} {2011})},\ \Eprint
  {http://arxiv.org/abs/1101.2599} {arXiv:1101.2599 [hep-ph]} \BibitemShut
  {NoStop}%
\bibitem [{\citenamefont {Fischer}\ and\ \citenamefont
  {Sjöstrand}(2017)}]{Fischer:2016zzs}%
  \BibitemOpen
  \bibfield  {author} {\bibinfo {author} {\bibfnamefont {N.}~\bibnamefont
  {Fischer}}\ and\ \bibinfo {author} {\bibfnamefont {T.}~\bibnamefont
  {Sjöstrand}},\ }\href {\doibase 10.1007/JHEP01(2017)140} {\bibfield
  {journal} {\bibinfo  {journal} {JHEP}\ }\textbf {\bibinfo {volume} {01}},\
  \bibinfo {pages} {140} (\bibinfo {year} {2017})},\ \Eprint
  {http://arxiv.org/abs/1610.09818} {arXiv:1610.09818 [hep-ph]} \BibitemShut
  {NoStop}%
\bibitem [{\citenamefont {Sjöstrand}(2019)}]{Sjostrand:2018xcd}%
  \BibitemOpen
  \bibfield  {author} {\bibinfo {author} {\bibfnamefont {T.}~\bibnamefont
  {Sjöstrand}},\ }\bibfield  {booktitle} {\emph {\bibinfo {booktitle}
  {{Proceedings, 27th International Conference on Ultrarelativistic
  Nucleus-Nucleus Collisions (Quark Matter 2018): Venice, Italy, May 14-19,
  2018}}},\ }\href {\doibase 10.1016/j.nuclphysa.2018.11.010} {\bibfield
  {journal} {\bibinfo  {journal} {Nucl. Phys.}\ }\textbf {\bibinfo {volume}
  {A982}},\ \bibinfo {pages} {43} (\bibinfo {year} {2019})},\ \Eprint
  {http://arxiv.org/abs/1808.03117} {arXiv:1808.03117 [hep-ph]} \BibitemShut
  {NoStop}%
\bibitem [{\citenamefont {Heinz}\ and\ \citenamefont
  {Snellings}(2013)}]{Heinz:2013th}%
  \BibitemOpen
  \bibfield  {author} {\bibinfo {author} {\bibfnamefont {U.}~\bibnamefont
  {Heinz}}\ and\ \bibinfo {author} {\bibfnamefont {R.}~\bibnamefont
  {Snellings}},\ }\href {\doibase 10.1146/annurev-nucl-102212-170540}
  {\bibfield  {journal} {\bibinfo  {journal} {Ann. Rev. Nucl. Part. Sci.}\
  }\textbf {\bibinfo {volume} {63}},\ \bibinfo {pages} {123} (\bibinfo {year}
  {2013})},\ \Eprint {http://arxiv.org/abs/1301.2826} {arXiv:1301.2826
  [nucl-th]} \BibitemShut {NoStop}%
\bibitem [{\citenamefont {Teaney}(2010)}]{Teaney:2009qa}%
  \BibitemOpen
  \bibfield  {author} {\bibinfo {author} {\bibfnamefont {D.~A.}\ \bibnamefont
  {Teaney}},\ }in\ \href {\doibase 10.1142/9789814293297_0004} {\emph {\bibinfo
  {booktitle} {Quark-gluon plasma 4}}},\ \bibinfo {editor} {edited by\ \bibinfo
  {editor} {\bibfnamefont {R.~C.}\ \bibnamefont {Hwa}}\ and\ \bibinfo {editor}
  {\bibfnamefont {X.-N.}\ \bibnamefont {Wang}}}\ (\bibinfo {year} {2010})\ pp.\
  \bibinfo {pages} {207--266},\ \Eprint {http://arxiv.org/abs/0905.2433}
  {arXiv:0905.2433 [nucl-th]} \BibitemShut {NoStop}%
\bibitem [{\citenamefont {Luzum}\ and\ \citenamefont
  {Petersen}(2014)}]{Luzum:2013yya}%
  \BibitemOpen
  \bibfield  {author} {\bibinfo {author} {\bibfnamefont {M.}~\bibnamefont
  {Luzum}}\ and\ \bibinfo {author} {\bibfnamefont {H.}~\bibnamefont
  {Petersen}},\ }\href {\doibase 10.1088/0954-3899/41/6/063102} {\bibfield
  {journal} {\bibinfo  {journal} {J. Phys.}\ }\textbf {\bibinfo {volume}
  {G41}},\ \bibinfo {pages} {063102} (\bibinfo {year} {2014})},\ \Eprint
  {http://arxiv.org/abs/1312.5503} {arXiv:1312.5503 [nucl-th]} \BibitemShut
  {NoStop}%
\bibitem [{\citenamefont {Gale}\ \emph {et~al.}(2013)\citenamefont {Gale},
  \citenamefont {Jeon},\ and\ \citenamefont {Schenke}}]{Gale:2013da}%
  \BibitemOpen
  \bibfield  {author} {\bibinfo {author} {\bibfnamefont {C.}~\bibnamefont
  {Gale}}, \bibinfo {author} {\bibfnamefont {S.}~\bibnamefont {Jeon}}, \ and\
  \bibinfo {author} {\bibfnamefont {B.}~\bibnamefont {Schenke}},\ }\href
  {\doibase 10.1142/S0217751X13400113} {\bibfield  {journal} {\bibinfo
  {journal} {Int. J. Mod. Phys.}\ }\textbf {\bibinfo {volume} {A28}},\ \bibinfo
  {pages} {1340011} (\bibinfo {year} {2013})},\ \Eprint
  {http://arxiv.org/abs/1301.5893} {arXiv:1301.5893 [nucl-th]} \BibitemShut
  {NoStop}%
\bibitem [{\citenamefont {Derradi~de Souza}\ \emph {et~al.}(2016)\citenamefont
  {Derradi~de Souza}, \citenamefont {Koide},\ and\ \citenamefont
  {Kodama}}]{deSouza:2015ena}%
  \BibitemOpen
  \bibfield  {author} {\bibinfo {author} {\bibfnamefont {R.}~\bibnamefont
  {Derradi~de Souza}}, \bibinfo {author} {\bibfnamefont {T.}~\bibnamefont
  {Koide}}, \ and\ \bibinfo {author} {\bibfnamefont {T.}~\bibnamefont
  {Kodama}},\ }\href {\doibase 10.1016/j.ppnp.2015.09.002} {\bibfield
  {journal} {\bibinfo  {journal} {Prog. Part. Nucl. Phys.}\ }\textbf {\bibinfo
  {volume} {86}},\ \bibinfo {pages} {35} (\bibinfo {year} {2016})},\ \Eprint
  {http://arxiv.org/abs/1506.03863} {arXiv:1506.03863 [nucl-th]} \BibitemShut
  {NoStop}%
\bibitem [{\citenamefont {Braun-Munzinger}\ \emph {et~al.}(2003)\citenamefont
  {Braun-Munzinger}, \citenamefont {Redlich},\ and\ \citenamefont
  {Stachel}}]{BraunMunzinger:2003zd}%
  \BibitemOpen
  \bibfield  {author} {\bibinfo {author} {\bibfnamefont {P.}~\bibnamefont
  {Braun-Munzinger}}, \bibinfo {author} {\bibfnamefont {K.}~\bibnamefont
  {Redlich}}, \ and\ \bibinfo {author} {\bibfnamefont {J.}~\bibnamefont
  {Stachel}},\ }\href {\doibase 10.1142/9789812795533_0008} {\ ,\ \bibinfo
  {pages} {491} (\bibinfo {year} {2003})},\ \Eprint
  {http://arxiv.org/abs/nucl-th/0304013} {arXiv:nucl-th/0304013 [nucl-th]}
  \BibitemShut {NoStop}%
\bibitem [{\citenamefont {Cleymans}\ \emph {et~al.}(2006)\citenamefont
  {Cleymans}, \citenamefont {Kraus}, \citenamefont {Oeschler}, \citenamefont
  {Redlich},\ and\ \citenamefont {Wheaton}}]{Cleymans:2006xj}%
  \BibitemOpen
  \bibfield  {author} {\bibinfo {author} {\bibfnamefont {J.}~\bibnamefont
  {Cleymans}}, \bibinfo {author} {\bibfnamefont {I.}~\bibnamefont {Kraus}},
  \bibinfo {author} {\bibfnamefont {H.}~\bibnamefont {Oeschler}}, \bibinfo
  {author} {\bibfnamefont {K.}~\bibnamefont {Redlich}}, \ and\ \bibinfo
  {author} {\bibfnamefont {S.}~\bibnamefont {Wheaton}},\ }\href {\doibase
  10.1103/PhysRevC.74.034903} {\bibfield  {journal} {\bibinfo  {journal} {Phys.
  Rev.}\ }\textbf {\bibinfo {volume} {C74}},\ \bibinfo {pages} {034903}
  (\bibinfo {year} {2006})},\ \Eprint {http://arxiv.org/abs/hep-ph/0604237}
  {arXiv:hep-ph/0604237 [hep-ph]} \BibitemShut {NoStop}%
\bibitem [{\citenamefont {Andronic}\ \emph {et~al.}(2009)\citenamefont
  {Andronic}, \citenamefont {Braun-Munzinger},\ and\ \citenamefont
  {Stachel}}]{Andronic:2008gu}%
  \BibitemOpen
  \bibfield  {author} {\bibinfo {author} {\bibfnamefont {A.}~\bibnamefont
  {Andronic}}, \bibinfo {author} {\bibfnamefont {P.}~\bibnamefont
  {Braun-Munzinger}}, \ and\ \bibinfo {author} {\bibfnamefont {J.}~\bibnamefont
  {Stachel}},\ }\href {\doibase 10.1016/j.physletb.2009.02.014,
  10.1016/j.physletb.2009.06.021} {\bibfield  {journal} {\bibinfo  {journal}
  {Phys. Lett.}\ }\textbf {\bibinfo {volume} {B673}},\ \bibinfo {pages} {142}
  (\bibinfo {year} {2009})},\ \bibinfo {note} {[Erratum: Phys.
  Lett.B678,516(2009)]},\ \Eprint {http://arxiv.org/abs/0812.1186}
  {arXiv:0812.1186 [nucl-th]} \BibitemShut {NoStop}%
\bibitem [{\citenamefont {Andronic}\ \emph {et~al.}(2018)\citenamefont
  {Andronic}, \citenamefont {Braun-Munzinger}, \citenamefont {Redlich},\ and\
  \citenamefont {Stachel}}]{Andronic:2017pug}%
  \BibitemOpen
  \bibfield  {author} {\bibinfo {author} {\bibfnamefont {A.}~\bibnamefont
  {Andronic}}, \bibinfo {author} {\bibfnamefont {P.}~\bibnamefont
  {Braun-Munzinger}}, \bibinfo {author} {\bibfnamefont {K.}~\bibnamefont
  {Redlich}}, \ and\ \bibinfo {author} {\bibfnamefont {J.}~\bibnamefont
  {Stachel}},\ }\href {\doibase 10.1038/s41586-018-0491-6} {\bibfield
  {journal} {\bibinfo  {journal} {Nature}\ }\textbf {\bibinfo {volume} {561}},\
  \bibinfo {pages} {321} (\bibinfo {year} {2018})},\ \Eprint
  {http://arxiv.org/abs/1710.09425} {arXiv:1710.09425 [nucl-th]} \BibitemShut
  {NoStop}%
\bibitem [{\citenamefont {Berges}\ \emph
  {et~al.}(2014{\natexlab{a}})\citenamefont {Berges}, \citenamefont
  {Boguslavski}, \citenamefont {Schlichting},\ and\ \citenamefont
  {Venugopalan}}]{Berges:2013fga}%
  \BibitemOpen
  \bibfield  {author} {\bibinfo {author} {\bibfnamefont {J.}~\bibnamefont
  {Berges}}, \bibinfo {author} {\bibfnamefont {K.}~\bibnamefont {Boguslavski}},
  \bibinfo {author} {\bibfnamefont {S.}~\bibnamefont {Schlichting}}, \ and\
  \bibinfo {author} {\bibfnamefont {R.}~\bibnamefont {Venugopalan}},\ }\href
  {\doibase 10.1103/PhysRevD.89.114007} {\bibfield  {journal} {\bibinfo
  {journal} {Phys. Rev.}\ }\textbf {\bibinfo {volume} {D89}},\ \bibinfo {pages}
  {114007} (\bibinfo {year} {2014}{\natexlab{a}})},\ \Eprint
  {http://arxiv.org/abs/1311.3005} {arXiv:1311.3005 [hep-ph]} \BibitemShut
  {NoStop}%
\bibitem [{\citenamefont {Berges}\ \emph
  {et~al.}(2014{\natexlab{b}})\citenamefont {Berges}, \citenamefont
  {Boguslavski}, \citenamefont {Schlichting},\ and\ \citenamefont
  {Venugopalan}}]{Berges:2013eia}%
  \BibitemOpen
  \bibfield  {author} {\bibinfo {author} {\bibfnamefont {J.}~\bibnamefont
  {Berges}}, \bibinfo {author} {\bibfnamefont {K.}~\bibnamefont {Boguslavski}},
  \bibinfo {author} {\bibfnamefont {S.}~\bibnamefont {Schlichting}}, \ and\
  \bibinfo {author} {\bibfnamefont {R.}~\bibnamefont {Venugopalan}},\ }\href
  {\doibase 10.1103/PhysRevD.89.074011} {\bibfield  {journal} {\bibinfo
  {journal} {Phys. Rev.}\ }\textbf {\bibinfo {volume} {D89}},\ \bibinfo {pages}
  {074011} (\bibinfo {year} {2014}{\natexlab{b}})},\ \Eprint
  {http://arxiv.org/abs/1303.5650} {arXiv:1303.5650 [hep-ph]} \BibitemShut
  {NoStop}%
\bibitem [{\citenamefont {Kurkela}\ and\ \citenamefont
  {Lu}(2014)}]{Kurkela:2014tea}%
  \BibitemOpen
  \bibfield  {author} {\bibinfo {author} {\bibfnamefont {A.}~\bibnamefont
  {Kurkela}}\ and\ \bibinfo {author} {\bibfnamefont {E.}~\bibnamefont {Lu}},\
  }\href {\doibase 10.1103/PhysRevLett.113.182301} {\bibfield  {journal}
  {\bibinfo  {journal} {Phys. Rev. Lett.}\ }\textbf {\bibinfo {volume} {113}},\
  \bibinfo {pages} {182301} (\bibinfo {year} {2014})},\ \Eprint
  {http://arxiv.org/abs/1405.6318} {arXiv:1405.6318 [hep-ph]} \BibitemShut
  {NoStop}%
\bibitem [{\citenamefont {Kurkela}\ and\ \citenamefont
  {Zhu}(2015)}]{Kurkela:2015qoa}%
  \BibitemOpen
  \bibfield  {author} {\bibinfo {author} {\bibfnamefont {A.}~\bibnamefont
  {Kurkela}}\ and\ \bibinfo {author} {\bibfnamefont {Y.}~\bibnamefont {Zhu}},\
  }\href {\doibase 10.1103/PhysRevLett.115.182301} {\bibfield  {journal}
  {\bibinfo  {journal} {Phys. Rev. Lett.}\ }\textbf {\bibinfo {volume} {115}},\
  \bibinfo {pages} {182301} (\bibinfo {year} {2015})},\ \Eprint
  {http://arxiv.org/abs/1506.06647} {arXiv:1506.06647 [hep-ph]} \BibitemShut
  {NoStop}%
\bibitem [{\citenamefont {Keegan}\ \emph
  {et~al.}(2016{\natexlab{a}})\citenamefont {Keegan}, \citenamefont {Kurkela},
  \citenamefont {Romatschke}, \citenamefont {van~der Schee},\ and\
  \citenamefont {Zhu}}]{Keegan:2015avk}%
  \BibitemOpen
  \bibfield  {author} {\bibinfo {author} {\bibfnamefont {L.}~\bibnamefont
  {Keegan}}, \bibinfo {author} {\bibfnamefont {A.}~\bibnamefont {Kurkela}},
  \bibinfo {author} {\bibfnamefont {P.}~\bibnamefont {Romatschke}}, \bibinfo
  {author} {\bibfnamefont {W.}~\bibnamefont {van~der Schee}}, \ and\ \bibinfo
  {author} {\bibfnamefont {Y.}~\bibnamefont {Zhu}},\ }\href {\doibase
  10.1007/JHEP04(2016)031} {\bibfield  {journal} {\bibinfo  {journal} {JHEP}\
  }\textbf {\bibinfo {volume} {04}},\ \bibinfo {pages} {031} (\bibinfo {year}
  {2016}{\natexlab{a}})},\ \Eprint {http://arxiv.org/abs/1512.05347}
  {arXiv:1512.05347 [hep-th]} \BibitemShut {NoStop}%
\bibitem [{\citenamefont {Keegan}\ \emph
  {et~al.}(2016{\natexlab{b}})\citenamefont {Keegan}, \citenamefont {Kurkela},
  \citenamefont {Mazeliauskas},\ and\ \citenamefont {Teaney}}]{Keegan:2016cpi}%
  \BibitemOpen
  \bibfield  {author} {\bibinfo {author} {\bibfnamefont {L.}~\bibnamefont
  {Keegan}}, \bibinfo {author} {\bibfnamefont {A.}~\bibnamefont {Kurkela}},
  \bibinfo {author} {\bibfnamefont {A.}~\bibnamefont {Mazeliauskas}}, \ and\
  \bibinfo {author} {\bibfnamefont {D.}~\bibnamefont {Teaney}},\ }\href
  {\doibase 10.1007/JHEP08(2016)171} {\bibfield  {journal} {\bibinfo  {journal}
  {JHEP}\ }\textbf {\bibinfo {volume} {08}},\ \bibinfo {pages} {171} (\bibinfo
  {year} {2016}{\natexlab{b}})},\ \Eprint {http://arxiv.org/abs/1605.04287}
  {arXiv:1605.04287 [hep-ph]} \BibitemShut {NoStop}%
\bibitem [{\citenamefont {Heller}\ and\ \citenamefont
  {Spalinski}(2015)}]{Heller:2015dha}%
  \BibitemOpen
  \bibfield  {author} {\bibinfo {author} {\bibfnamefont {M.~P.}\ \bibnamefont
  {Heller}}\ and\ \bibinfo {author} {\bibfnamefont {M.}~\bibnamefont
  {Spalinski}},\ }\href {\doibase 10.1103/PhysRevLett.115.072501} {\bibfield
  {journal} {\bibinfo  {journal} {Phys. Rev. Lett.}\ }\textbf {\bibinfo
  {volume} {115}},\ \bibinfo {pages} {072501} (\bibinfo {year} {2015})},\
  \Eprint {http://arxiv.org/abs/1503.07514} {arXiv:1503.07514 [hep-th]}
  \BibitemShut {NoStop}%
\bibitem [{\citenamefont {Heller}\ \emph {et~al.}(2018)\citenamefont {Heller},
  \citenamefont {Kurkela}, \citenamefont {Spaliński},\ and\ \citenamefont
  {Svensson}}]{Heller:2016rtz}%
  \BibitemOpen
  \bibfield  {author} {\bibinfo {author} {\bibfnamefont {M.~P.}\ \bibnamefont
  {Heller}}, \bibinfo {author} {\bibfnamefont {A.}~\bibnamefont {Kurkela}},
  \bibinfo {author} {\bibfnamefont {M.}~\bibnamefont {Spaliński}}, \ and\
  \bibinfo {author} {\bibfnamefont {V.}~\bibnamefont {Svensson}},\ }\href
  {\doibase 10.1103/PhysRevD.97.091503} {\bibfield  {journal} {\bibinfo
  {journal} {Phys. Rev.}\ }\textbf {\bibinfo {volume} {D97}},\ \bibinfo {pages}
  {091503} (\bibinfo {year} {2018})},\ \Eprint
  {http://arxiv.org/abs/1609.04803} {arXiv:1609.04803 [nucl-th]} \BibitemShut
  {NoStop}%
\bibitem [{\citenamefont {Strickland}\ \emph {et~al.}(2018)\citenamefont
  {Strickland}, \citenamefont {Noronha},\ and\ \citenamefont
  {Denicol}}]{Strickland:2017kux}%
  \BibitemOpen
  \bibfield  {author} {\bibinfo {author} {\bibfnamefont {M.}~\bibnamefont
  {Strickland}}, \bibinfo {author} {\bibfnamefont {J.}~\bibnamefont {Noronha}},
  \ and\ \bibinfo {author} {\bibfnamefont {G.}~\bibnamefont {Denicol}},\ }\href
  {\doibase 10.1103/PhysRevD.97.036020} {\bibfield  {journal} {\bibinfo
  {journal} {Phys. Rev.}\ }\textbf {\bibinfo {volume} {D97}},\ \bibinfo {pages}
  {036020} (\bibinfo {year} {2018})},\ \Eprint
  {http://arxiv.org/abs/1709.06644} {arXiv:1709.06644 [nucl-th]} \BibitemShut
  {NoStop}%
\bibitem [{\citenamefont {Behtash}\ \emph {et~al.}(2018)\citenamefont
  {Behtash}, \citenamefont {Cruz-Camacho},\ and\ \citenamefont
  {Martinez}}]{Behtash:2017wqg}%
  \BibitemOpen
  \bibfield  {author} {\bibinfo {author} {\bibfnamefont {A.}~\bibnamefont
  {Behtash}}, \bibinfo {author} {\bibfnamefont {C.~N.}\ \bibnamefont
  {Cruz-Camacho}}, \ and\ \bibinfo {author} {\bibfnamefont {M.}~\bibnamefont
  {Martinez}},\ }\href {\doibase 10.1103/PhysRevD.97.044041} {\bibfield
  {journal} {\bibinfo  {journal} {Phys. Rev.}\ }\textbf {\bibinfo {volume}
  {D97}},\ \bibinfo {pages} {044041} (\bibinfo {year} {2018})},\ \Eprint
  {http://arxiv.org/abs/1711.01745} {arXiv:1711.01745 [hep-th]} \BibitemShut
  {NoStop}%
\bibitem [{\citenamefont {Romatschke}(2018)}]{Romatschke:2017vte}%
  \BibitemOpen
  \bibfield  {author} {\bibinfo {author} {\bibfnamefont {P.}~\bibnamefont
  {Romatschke}},\ }\href {\doibase 10.1103/PhysRevLett.120.012301} {\bibfield
  {journal} {\bibinfo  {journal} {Phys. Rev. Lett.}\ }\textbf {\bibinfo
  {volume} {120}},\ \bibinfo {pages} {012301} (\bibinfo {year} {2018})},\
  \Eprint {http://arxiv.org/abs/1704.08699} {arXiv:1704.08699 [hep-th]}
  \BibitemShut {NoStop}%
\bibitem [{\citenamefont {Arnold}\ \emph
  {et~al.}(2003{\natexlab{a}})\citenamefont {Arnold}, \citenamefont {Moore},\
  and\ \citenamefont {Yaffe}}]{Arnold:2003zc}%
  \BibitemOpen
  \bibfield  {author} {\bibinfo {author} {\bibfnamefont {P.~B.}\ \bibnamefont
  {Arnold}}, \bibinfo {author} {\bibfnamefont {G.~D.}\ \bibnamefont {Moore}}, \
  and\ \bibinfo {author} {\bibfnamefont {L.~G.}\ \bibnamefont {Yaffe}},\ }\href
  {\doibase 10.1088/1126-6708/2003/05/051} {\bibfield  {journal} {\bibinfo
  {journal} {JHEP}\ }\textbf {\bibinfo {volume} {05}},\ \bibinfo {pages} {051}
  (\bibinfo {year} {2003}{\natexlab{a}})},\ \Eprint
  {http://arxiv.org/abs/hep-ph/0302165} {arXiv:hep-ph/0302165 [hep-ph]}
  \BibitemShut {NoStop}%
\bibitem [{\citenamefont {Rafelski}\ and\ \citenamefont
  {Muller}(1982)}]{Rafelski:1982pu}%
  \BibitemOpen
  \bibfield  {author} {\bibinfo {author} {\bibfnamefont {J.}~\bibnamefont
  {Rafelski}}\ and\ \bibinfo {author} {\bibfnamefont {B.}~\bibnamefont
  {Muller}},\ }\href {\doibase 10.1103/PhysRevLett.48.1066,
  10.1103/PhysRevLett.56.2334} {\bibfield  {journal} {\bibinfo  {journal}
  {Phys. Rev. Lett.}\ }\textbf {\bibinfo {volume} {48}},\ \bibinfo {pages}
  {1066} (\bibinfo {year} {1982})},\ \bibinfo {note} {[Erratum: Phys. Rev.
  Lett.56,2334(1986)]}\BibitemShut {NoStop}%
\bibitem [{\citenamefont {Shuryak}(1992)}]{Shuryak:1992wc}%
  \BibitemOpen
  \bibfield  {author} {\bibinfo {author} {\bibfnamefont {E.~V.}\ \bibnamefont
  {Shuryak}},\ }\href {\doibase 10.1103/PhysRevLett.68.3270} {\bibfield
  {journal} {\bibinfo  {journal} {Phys. Rev. Lett.}\ }\textbf {\bibinfo
  {volume} {68}},\ \bibinfo {pages} {3270} (\bibinfo {year}
  {1992})}\BibitemShut {NoStop}%
\bibitem [{\citenamefont {Gelfand}\ \emph {et~al.}(2016)\citenamefont
  {Gelfand}, \citenamefont {Hebenstreit},\ and\ \citenamefont
  {Berges}}]{Gelfand:2016prm}%
  \BibitemOpen
  \bibfield  {author} {\bibinfo {author} {\bibfnamefont {D.}~\bibnamefont
  {Gelfand}}, \bibinfo {author} {\bibfnamefont {F.}~\bibnamefont
  {Hebenstreit}}, \ and\ \bibinfo {author} {\bibfnamefont {J.}~\bibnamefont
  {Berges}},\ }\href {\doibase 10.1103/PhysRevD.93.085001} {\bibfield
  {journal} {\bibinfo  {journal} {Phys. Rev.}\ }\textbf {\bibinfo {volume}
  {D93}},\ \bibinfo {pages} {085001} (\bibinfo {year} {2016})},\ \Eprint
  {http://arxiv.org/abs/1601.03576} {arXiv:1601.03576 [hep-ph]} \BibitemShut
  {NoStop}%
\bibitem [{\citenamefont {Tanji}\ and\ \citenamefont
  {Berges}(2018)}]{Tanji:2017xiw}%
  \BibitemOpen
  \bibfield  {author} {\bibinfo {author} {\bibfnamefont {N.}~\bibnamefont
  {Tanji}}\ and\ \bibinfo {author} {\bibfnamefont {J.}~\bibnamefont {Berges}},\
  }\href {\doibase 10.1103/PhysRevD.97.034013} {\bibfield  {journal} {\bibinfo
  {journal} {Phys. Rev.}\ }\textbf {\bibinfo {volume} {D97}},\ \bibinfo {pages}
  {034013} (\bibinfo {year} {2018})},\ \Eprint
  {http://arxiv.org/abs/1711.03445} {arXiv:1711.03445 [hep-ph]} \BibitemShut
  {NoStop}%
\bibitem [{\citenamefont {Biro}\ \emph {et~al.}(1993)\citenamefont {Biro},
  \citenamefont {van Doorn}, \citenamefont {Muller}, \citenamefont {Thoma},\
  and\ \citenamefont {Wang}}]{Biro:1993qt}%
  \BibitemOpen
  \bibfield  {author} {\bibinfo {author} {\bibfnamefont {T.~S.}\ \bibnamefont
  {Biro}}, \bibinfo {author} {\bibfnamefont {E.}~\bibnamefont {van Doorn}},
  \bibinfo {author} {\bibfnamefont {B.}~\bibnamefont {Muller}}, \bibinfo
  {author} {\bibfnamefont {M.~H.}\ \bibnamefont {Thoma}}, \ and\ \bibinfo
  {author} {\bibfnamefont {X.~N.}\ \bibnamefont {Wang}},\ }\href {\doibase
  10.1103/PhysRevC.48.1275} {\bibfield  {journal} {\bibinfo  {journal} {Phys.
  Rev.}\ }\textbf {\bibinfo {volume} {C48}},\ \bibinfo {pages} {1275} (\bibinfo
  {year} {1993})},\ \Eprint {http://arxiv.org/abs/nucl-th/9303004}
  {arXiv:nucl-th/9303004 [nucl-th]} \BibitemShut {NoStop}%
\bibitem [{\citenamefont {Elliott}\ and\ \citenamefont
  {Rischke}(2000)}]{Elliott:1999uz}%
  \BibitemOpen
  \bibfield  {author} {\bibinfo {author} {\bibfnamefont {D.~M.}\ \bibnamefont
  {Elliott}}\ and\ \bibinfo {author} {\bibfnamefont {D.~H.}\ \bibnamefont
  {Rischke}},\ }\href {\doibase 10.1016/S0375-9474(99)00840-4} {\bibfield
  {journal} {\bibinfo  {journal} {Nucl. Phys.}\ }\textbf {\bibinfo {volume}
  {A671}},\ \bibinfo {pages} {583} (\bibinfo {year} {2000})},\ \Eprint
  {http://arxiv.org/abs/nucl-th/9908004} {arXiv:nucl-th/9908004 [nucl-th]}
  \BibitemShut {NoStop}%
\bibitem [{\citenamefont {Geiger}\ and\ \citenamefont
  {Muller}(1992)}]{Geiger:1991nj}%
  \BibitemOpen
  \bibfield  {author} {\bibinfo {author} {\bibfnamefont {K.}~\bibnamefont
  {Geiger}}\ and\ \bibinfo {author} {\bibfnamefont {B.}~\bibnamefont
  {Muller}},\ }\href {\doibase 10.1016/0550-3213(92)90280-O} {\bibfield
  {journal} {\bibinfo  {journal} {Nucl. Phys.}\ }\textbf {\bibinfo {volume}
  {B369}},\ \bibinfo {pages} {600} (\bibinfo {year} {1992})}\BibitemShut
  {NoStop}%
\bibitem [{\citenamefont {Borchers}\ \emph {et~al.}(2000)\citenamefont
  {Borchers}, \citenamefont {Meyer}, \citenamefont {Gieseke}, \citenamefont
  {Martens},\ and\ \citenamefont {Noack}}]{Borchers:2000wf}%
  \BibitemOpen
  \bibfield  {author} {\bibinfo {author} {\bibfnamefont {V.}~\bibnamefont
  {Borchers}}, \bibinfo {author} {\bibfnamefont {J.}~\bibnamefont {Meyer}},
  \bibinfo {author} {\bibfnamefont {S.}~\bibnamefont {Gieseke}}, \bibinfo
  {author} {\bibfnamefont {G.}~\bibnamefont {Martens}}, \ and\ \bibinfo
  {author} {\bibfnamefont {C.~C.}\ \bibnamefont {Noack}},\ }\href {\doibase
  10.1103/PhysRevC.62.064903} {\bibfield  {journal} {\bibinfo  {journal} {Phys.
  Rev.}\ }\textbf {\bibinfo {volume} {C62}},\ \bibinfo {pages} {064903}
  (\bibinfo {year} {2000})},\ \Eprint {http://arxiv.org/abs/hep-ph/0006038}
  {arXiv:hep-ph/0006038 [hep-ph]} \BibitemShut {NoStop}%
\bibitem [{\citenamefont {Xu}\ and\ \citenamefont {Greiner}(2005)}]{Xu:2004mz}%
  \BibitemOpen
  \bibfield  {author} {\bibinfo {author} {\bibfnamefont {Z.}~\bibnamefont
  {Xu}}\ and\ \bibinfo {author} {\bibfnamefont {C.}~\bibnamefont {Greiner}},\
  }\href {\doibase 10.1103/PhysRevC.71.064901} {\bibfield  {journal} {\bibinfo
  {journal} {Phys. Rev.}\ }\textbf {\bibinfo {volume} {C71}},\ \bibinfo {pages}
  {064901} (\bibinfo {year} {2005})},\ \Eprint
  {http://arxiv.org/abs/hep-ph/0406278} {arXiv:hep-ph/0406278 [hep-ph]}
  \BibitemShut {NoStop}%
\bibitem [{\citenamefont {Blaizot}\ \emph {et~al.}(2014)\citenamefont
  {Blaizot}, \citenamefont {Wu},\ and\ \citenamefont {Yan}}]{Blaizot:2014jna}%
  \BibitemOpen
  \bibfield  {author} {\bibinfo {author} {\bibfnamefont {J.-P.}\ \bibnamefont
  {Blaizot}}, \bibinfo {author} {\bibfnamefont {B.}~\bibnamefont {Wu}}, \ and\
  \bibinfo {author} {\bibfnamefont {L.}~\bibnamefont {Yan}},\ }\href {\doibase
  10.1016/j.nuclphysa.2014.07.041} {\bibfield  {journal} {\bibinfo  {journal}
  {Nucl. Phys.}\ }\textbf {\bibinfo {volume} {A930}},\ \bibinfo {pages} {139}
  (\bibinfo {year} {2014})},\ \Eprint {http://arxiv.org/abs/1402.5049}
  {arXiv:1402.5049 [hep-ph]} \BibitemShut {NoStop}%
\bibitem [{\citenamefont {Ruggieri}\ \emph {et~al.}(2015)\citenamefont
  {Ruggieri}, \citenamefont {Plumari}, \citenamefont {Scardina},\ and\
  \citenamefont {Greco}}]{Ruggieri:2015tsa}%
  \BibitemOpen
  \bibfield  {author} {\bibinfo {author} {\bibfnamefont {M.}~\bibnamefont
  {Ruggieri}}, \bibinfo {author} {\bibfnamefont {S.}~\bibnamefont {Plumari}},
  \bibinfo {author} {\bibfnamefont {F.}~\bibnamefont {Scardina}}, \ and\
  \bibinfo {author} {\bibfnamefont {V.}~\bibnamefont {Greco}},\ }\href
  {\doibase 10.1016/j.nuclphysa.2015.07.004} {\bibfield  {journal} {\bibinfo
  {journal} {Nucl. Phys.}\ }\textbf {\bibinfo {volume} {A941}},\ \bibinfo
  {pages} {201} (\bibinfo {year} {2015})},\ \Eprint
  {http://arxiv.org/abs/1502.04596} {arXiv:1502.04596 [nucl-th]} \BibitemShut
  {NoStop}%
\bibitem [{\citenamefont {Arnold}\ \emph
  {et~al.}(2003{\natexlab{b}})\citenamefont {Arnold}, \citenamefont {Moore},\
  and\ \citenamefont {Yaffe}}]{Arnold:2002zm}%
  \BibitemOpen
  \bibfield  {author} {\bibinfo {author} {\bibfnamefont {P.~B.}\ \bibnamefont
  {Arnold}}, \bibinfo {author} {\bibfnamefont {G.~D.}\ \bibnamefont {Moore}}, \
  and\ \bibinfo {author} {\bibfnamefont {L.~G.}\ \bibnamefont {Yaffe}},\ }\href
  {\doibase 10.1088/1126-6708/2003/01/030} {\bibfield  {journal} {\bibinfo
  {journal} {JHEP}\ }\textbf {\bibinfo {volume} {01}},\ \bibinfo {pages} {030}
  (\bibinfo {year} {2003}{\natexlab{b}})},\ \Eprint
  {http://arxiv.org/abs/hep-ph/0209353} {arXiv:hep-ph/0209353 [hep-ph]}
  \BibitemShut {NoStop}%
\bibitem [{\citenamefont {Lappi}\ and\ \citenamefont
  {McLerran}(2006)}]{Lappi:2006fp}%
  \BibitemOpen
  \bibfield  {author} {\bibinfo {author} {\bibfnamefont {T.}~\bibnamefont
  {Lappi}}\ and\ \bibinfo {author} {\bibfnamefont {L.}~\bibnamefont
  {McLerran}},\ }\href {\doibase 10.1016/j.nuclphysa.2006.04.001} {\bibfield
  {journal} {\bibinfo  {journal} {Nucl. Phys.}\ }\textbf {\bibinfo {volume}
  {A772}},\ \bibinfo {pages} {200} (\bibinfo {year} {2006})},\ \Eprint
  {http://arxiv.org/abs/hep-ph/0602189} {arXiv:hep-ph/0602189 [hep-ph]}
  \BibitemShut {NoStop}%
\bibitem [{\citenamefont {Gelis}\ \emph {et~al.}(2010)\citenamefont {Gelis},
  \citenamefont {Iancu}, \citenamefont {Jalilian-Marian},\ and\ \citenamefont
  {Venugopalan}}]{Gelis:2010nm}%
  \BibitemOpen
  \bibfield  {author} {\bibinfo {author} {\bibfnamefont {F.}~\bibnamefont
  {Gelis}}, \bibinfo {author} {\bibfnamefont {E.}~\bibnamefont {Iancu}},
  \bibinfo {author} {\bibfnamefont {J.}~\bibnamefont {Jalilian-Marian}}, \ and\
  \bibinfo {author} {\bibfnamefont {R.}~\bibnamefont {Venugopalan}},\ }\href
  {\doibase 10.1146/annurev.nucl.010909.083629} {\bibfield  {journal} {\bibinfo
   {journal} {Ann. Rev. Nucl. Part. Sci.}\ }\textbf {\bibinfo {volume} {60}},\
  \bibinfo {pages} {463} (\bibinfo {year} {2010})},\ \Eprint
  {http://arxiv.org/abs/1002.0333} {arXiv:1002.0333 [hep-ph]} \BibitemShut
  {NoStop}%
\bibitem [{\citenamefont {Lappi}(2011)}]{Lappi:2011ju}%
  \BibitemOpen
  \bibfield  {author} {\bibinfo {author} {\bibfnamefont {T.}~\bibnamefont
  {Lappi}},\ }\href {\doibase 10.1016/j.physletb.2011.08.011} {\bibfield
  {journal} {\bibinfo  {journal} {Phys. Lett.}\ }\textbf {\bibinfo {volume}
  {B703}},\ \bibinfo {pages} {325} (\bibinfo {year} {2011})},\ \Eprint
  {http://arxiv.org/abs/1105.5511} {arXiv:1105.5511 [hep-ph]} \BibitemShut
  {NoStop}%
\bibitem [{\citenamefont {Braaten}\ and\ \citenamefont
  {Pisarski}(1990)}]{Braaten:1989mz}%
  \BibitemOpen
  \bibfield  {author} {\bibinfo {author} {\bibfnamefont {E.}~\bibnamefont
  {Braaten}}\ and\ \bibinfo {author} {\bibfnamefont {R.~D.}\ \bibnamefont
  {Pisarski}},\ }\href {\doibase 10.1016/0550-3213(90)90508-B} {\bibfield
  {journal} {\bibinfo  {journal} {Nucl. Phys.}\ }\textbf {\bibinfo {volume}
  {B337}},\ \bibinfo {pages} {569} (\bibinfo {year} {1990})}\BibitemShut
  {NoStop}%
\bibitem [{\citenamefont {Landau}\ and\ \citenamefont
  {Pomeranchuk}(1953{\natexlab{a}})}]{Landau:1953gr}%
  \BibitemOpen
  \bibfield  {author} {\bibinfo {author} {\bibfnamefont {L.~D.}\ \bibnamefont
  {Landau}}\ and\ \bibinfo {author} {\bibfnamefont {I.}~\bibnamefont
  {Pomeranchuk}},\ }\href@noop {} {\bibfield  {journal} {\bibinfo  {journal}
  {Dokl. Akad. Nauk Ser. Fiz.}\ }\textbf {\bibinfo {volume} {92}},\ \bibinfo
  {pages} {735} (\bibinfo {year} {1953}{\natexlab{a}})}\BibitemShut {NoStop}%
\bibitem [{\citenamefont {Landau}\ and\ \citenamefont
  {Pomeranchuk}(1953{\natexlab{b}})}]{Landau:1953um}%
  \BibitemOpen
  \bibfield  {author} {\bibinfo {author} {\bibfnamefont {L.~D.}\ \bibnamefont
  {Landau}}\ and\ \bibinfo {author} {\bibfnamefont {I.}~\bibnamefont
  {Pomeranchuk}},\ }\href@noop {} {\bibfield  {journal} {\bibinfo  {journal}
  {Dokl. Akad. Nauk Ser. Fiz.}\ }\textbf {\bibinfo {volume} {92}},\ \bibinfo
  {pages} {535} (\bibinfo {year} {1953}{\natexlab{b}})}\BibitemShut {NoStop}%
\bibitem [{\citenamefont {Migdal}(1956)}]{Migdal:1956tc}%
  \BibitemOpen
  \bibfield  {author} {\bibinfo {author} {\bibfnamefont {A.~B.}\ \bibnamefont
  {Migdal}},\ }\href {\doibase 10.1103/PhysRev.103.1811} {\bibfield  {journal}
  {\bibinfo  {journal} {Phys. Rev.}\ }\textbf {\bibinfo {volume} {103}},\
  \bibinfo {pages} {1811} (\bibinfo {year} {1956})}\BibitemShut {NoStop}%
\bibitem [{\citenamefont {Migdal}(1955)}]{Migdal:1955nv}%
  \BibitemOpen
  \bibfield  {author} {\bibinfo {author} {\bibfnamefont {A.~B.}\ \bibnamefont
  {Migdal}},\ }\href@noop {} {\bibfield  {journal} {\bibinfo  {journal} {Dokl.
  Akad. Nauk Ser. Fiz.}\ }\textbf {\bibinfo {volume} {105}},\ \bibinfo {pages}
  {77} (\bibinfo {year} {1955})}\BibitemShut {NoStop}%
\bibitem [{Note1()}]{Note1}%
  \BibitemOpen
  \bibinfo {note} {Anisotropic systems suffer from the presence of unstable
  plasma modes~\cite {Mrowczynski:1988dz,
  Mrowczynski:1993qm,Mrowczynski:2000ed}, which could change the kinetic
  dynamics~\cite {Kurkela:2011ub,Kurkela:2011ti}. However detailed 3+1D
  classical-statistical YM simulations found no effects of plasma instabilities
  beyond very early times~\cite {Berges:2013fga,Berges:2013eia}. Therefore we
  will use isotropic approximations, which remove the unstable modes from the
  kinetic description. Note that there are no unstable fermionic modes~\cite
  {Mrowczynski:2001az,Schenke:2006fz}}\BibitemShut {NoStop}%
\bibitem [{\citenamefont {Kurkela}\ and\ \citenamefont
  {Mazeliauskas}(2019)}]{Kurkela:2018oqw}%
  \BibitemOpen
  \bibfield  {author} {\bibinfo {author} {\bibfnamefont {A.}~\bibnamefont
  {Kurkela}}\ and\ \bibinfo {author} {\bibfnamefont {A.}~\bibnamefont
  {Mazeliauskas}},\ }\href {\doibase 10.1103/PhysRevD.99.054018} {\bibfield
  {journal} {\bibinfo  {journal} {Phys. Rev.}\ }\textbf {\bibinfo {volume}
  {D99}},\ \bibinfo {pages} {054018} (\bibinfo {year} {2019})},\ \Eprint
  {http://arxiv.org/abs/1811.03068} {arXiv:1811.03068 [hep-ph]} \BibitemShut
  {NoStop}%
\bibitem [{\citenamefont {Bjorken}(1983)}]{Bjorken:1982qr}%
  \BibitemOpen
  \bibfield  {author} {\bibinfo {author} {\bibfnamefont {J.~D.}\ \bibnamefont
  {Bjorken}},\ }\href {\doibase 10.1103/PhysRevD.27.140} {\bibfield  {journal}
  {\bibinfo  {journal} {Phys. Rev.}\ }\textbf {\bibinfo {volume} {D27}},\
  \bibinfo {pages} {140} (\bibinfo {year} {1983})}\BibitemShut {NoStop}%
\bibitem [{\citenamefont {Abraao~York}\ \emph {et~al.}(2014)\citenamefont
  {Abraao~York}, \citenamefont {Kurkela}, \citenamefont {Lu},\ and\
  \citenamefont {Moore}}]{York:2014wja}%
  \BibitemOpen
  \bibfield  {author} {\bibinfo {author} {\bibfnamefont {M.~C.}\ \bibnamefont
  {Abraao~York}}, \bibinfo {author} {\bibfnamefont {A.}~\bibnamefont
  {Kurkela}}, \bibinfo {author} {\bibfnamefont {E.}~\bibnamefont {Lu}}, \ and\
  \bibinfo {author} {\bibfnamefont {G.~D.}\ \bibnamefont {Moore}},\ }\href
  {\doibase 10.1103/PhysRevD.89.074036} {\bibfield  {journal} {\bibinfo
  {journal} {Phys. Rev.}\ }\textbf {\bibinfo {volume} {D89}},\ \bibinfo {pages}
  {074036} (\bibinfo {year} {2014})},\ \Eprint {http://arxiv.org/abs/1401.3751}
  {arXiv:1401.3751 [hep-ph]} \BibitemShut {NoStop}%
\bibitem [{\citenamefont {Ghiglieri}\ \emph {et~al.}(2016)\citenamefont
  {Ghiglieri}, \citenamefont {Moore},\ and\ \citenamefont
  {Teaney}}]{Ghiglieri:2015ala}%
  \BibitemOpen
  \bibfield  {author} {\bibinfo {author} {\bibfnamefont {J.}~\bibnamefont
  {Ghiglieri}}, \bibinfo {author} {\bibfnamefont {G.~D.}\ \bibnamefont
  {Moore}}, \ and\ \bibinfo {author} {\bibfnamefont {D.}~\bibnamefont
  {Teaney}},\ }\href {\doibase 10.1007/JHEP03(2016)095} {\bibfield  {journal}
  {\bibinfo  {journal} {JHEP}\ }\textbf {\bibinfo {volume} {03}},\ \bibinfo
  {pages} {095} (\bibinfo {year} {2016})},\ \Eprint
  {http://arxiv.org/abs/1509.07773} {arXiv:1509.07773 [hep-ph]} \BibitemShut
  {NoStop}%
\bibitem [{\citenamefont {Teaney}()}]{teaney}%
  \BibitemOpen
  \bibfield  {author} {\bibinfo {author} {\bibfnamefont {D.}~\bibnamefont
  {Teaney}},\ }\href@noop {} {}\bibinfo {howpublished} {personal
  communication}\BibitemShut {NoStop}%
\bibitem [{\citenamefont {Aurenche}\ \emph {et~al.}(2002)\citenamefont
  {Aurenche}, \citenamefont {Gelis},\ and\ \citenamefont
  {Zaraket}}]{Aurenche:2002pd}%
  \BibitemOpen
  \bibfield  {author} {\bibinfo {author} {\bibfnamefont {P.}~\bibnamefont
  {Aurenche}}, \bibinfo {author} {\bibfnamefont {F.}~\bibnamefont {Gelis}}, \
  and\ \bibinfo {author} {\bibfnamefont {H.}~\bibnamefont {Zaraket}},\ }\href
  {\doibase 10.1088/1126-6708/2002/05/043} {\bibfield  {journal} {\bibinfo
  {journal} {JHEP}\ }\textbf {\bibinfo {volume} {05}},\ \bibinfo {pages} {043}
  (\bibinfo {year} {2002})},\ \Eprint {http://arxiv.org/abs/hep-ph/0204146}
  {arXiv:hep-ph/0204146 [hep-ph]} \BibitemShut {NoStop}%
\bibitem [{\citenamefont {Kurkela}\ \emph
  {et~al.}(2019{\natexlab{a}})\citenamefont {Kurkela}, \citenamefont
  {Mazeliauskas}, \citenamefont {Paquet}, \citenamefont {Schlichting},\ and\
  \citenamefont {Teaney}}]{Kurkela:2018vqr}%
  \BibitemOpen
  \bibfield  {author} {\bibinfo {author} {\bibfnamefont {A.}~\bibnamefont
  {Kurkela}}, \bibinfo {author} {\bibfnamefont {A.}~\bibnamefont
  {Mazeliauskas}}, \bibinfo {author} {\bibfnamefont {J.-F.}\ \bibnamefont
  {Paquet}}, \bibinfo {author} {\bibfnamefont {S.}~\bibnamefont {Schlichting}},
  \ and\ \bibinfo {author} {\bibfnamefont {D.}~\bibnamefont {Teaney}},\ }\href
  {\doibase 10.1103/PhysRevC.99.034910} {\bibfield  {journal} {\bibinfo
  {journal} {Phys. Rev.}\ }\textbf {\bibinfo {volume} {C99}},\ \bibinfo {pages}
  {034910} (\bibinfo {year} {2019}{\natexlab{a}})},\ \Eprint
  {http://arxiv.org/abs/1805.00961} {arXiv:1805.00961 [hep-ph]} \BibitemShut
  {NoStop}%
\bibitem [{\citenamefont {Kurkela}\ \emph
  {et~al.}(2019{\natexlab{b}})\citenamefont {Kurkela}, \citenamefont
  {Mazeliauskas}, \citenamefont {Paquet}, \citenamefont {Schlichting},\ and\
  \citenamefont {Teaney}}]{Kurkela:2018wud}%
  \BibitemOpen
  \bibfield  {author} {\bibinfo {author} {\bibfnamefont {A.}~\bibnamefont
  {Kurkela}}, \bibinfo {author} {\bibfnamefont {A.}~\bibnamefont
  {Mazeliauskas}}, \bibinfo {author} {\bibfnamefont {J.-F.}\ \bibnamefont
  {Paquet}}, \bibinfo {author} {\bibfnamefont {S.}~\bibnamefont {Schlichting}},
  \ and\ \bibinfo {author} {\bibfnamefont {D.}~\bibnamefont {Teaney}},\ }\href
  {\doibase 10.1103/PhysRevLett.122.122302} {\bibfield  {journal} {\bibinfo
  {journal} {Phys. Rev. Lett.}\ }\textbf {\bibinfo {volume} {122}},\ \bibinfo
  {pages} {122302} (\bibinfo {year} {2019}{\natexlab{b}})},\ \Eprint
  {http://arxiv.org/abs/1805.01604} {arXiv:1805.01604 [hep-ph]} \BibitemShut
  {NoStop}%
\bibitem [{\citenamefont {Tanabashi}\ \emph {et~al.}(2018)\citenamefont
  {Tanabashi} \emph {et~al.}}]{Tanabashi:2018oca}%
  \BibitemOpen
  \bibfield  {author} {\bibinfo {author} {\bibfnamefont {M.}~\bibnamefont
  {Tanabashi}} \emph {et~al.} (\bibinfo {collaboration} {Particle Data
  Group}),\ }\href {\doibase 10.1103/PhysRevD.98.030001} {\bibfield  {journal}
  {\bibinfo  {journal} {Phys. Rev.}\ }\textbf {\bibinfo {volume} {D98}},\
  \bibinfo {pages} {030001} (\bibinfo {year} {2018})}\BibitemShut {NoStop}%
\bibitem [{\citenamefont {Ghiglieri}\ \emph
  {et~al.}(2018{\natexlab{a}})\citenamefont {Ghiglieri}, \citenamefont
  {Moore},\ and\ \citenamefont {Teaney}}]{Ghiglieri:2018dib}%
  \BibitemOpen
  \bibfield  {author} {\bibinfo {author} {\bibfnamefont {J.}~\bibnamefont
  {Ghiglieri}}, \bibinfo {author} {\bibfnamefont {G.~D.}\ \bibnamefont
  {Moore}}, \ and\ \bibinfo {author} {\bibfnamefont {D.}~\bibnamefont
  {Teaney}},\ }\href {\doibase 10.1007/JHEP03(2018)179} {\bibfield  {journal}
  {\bibinfo  {journal} {JHEP}\ }\textbf {\bibinfo {volume} {03}},\ \bibinfo
  {pages} {179} (\bibinfo {year} {2018}{\natexlab{a}})},\ \Eprint
  {http://arxiv.org/abs/1802.09535} {arXiv:1802.09535 [hep-ph]} \BibitemShut
  {NoStop}%
\bibitem [{\citenamefont {Ghiglieri}\ \emph
  {et~al.}(2018{\natexlab{b}})\citenamefont {Ghiglieri}, \citenamefont
  {Moore},\ and\ \citenamefont {Teaney}}]{Ghiglieri:2018dgf}%
  \BibitemOpen
  \bibfield  {author} {\bibinfo {author} {\bibfnamefont {J.}~\bibnamefont
  {Ghiglieri}}, \bibinfo {author} {\bibfnamefont {G.~D.}\ \bibnamefont
  {Moore}}, \ and\ \bibinfo {author} {\bibfnamefont {D.}~\bibnamefont
  {Teaney}},\ }\href {\doibase 10.1103/PhysRevLett.121.052302} {\bibfield
  {journal} {\bibinfo  {journal} {Phys. Rev. Lett.}\ }\textbf {\bibinfo
  {volume} {121}},\ \bibinfo {pages} {052302} (\bibinfo {year}
  {2018}{\natexlab{b}})},\ \Eprint {http://arxiv.org/abs/1805.02663}
  {arXiv:1805.02663 [hep-ph]} \BibitemShut {NoStop}%
\bibitem [{Note2()}]{Note2}%
  \BibitemOpen
  \bibinfo {note} {We determine $\eta /s(\lambda )$ numerically from
  near-equilibrium pressure anisotropy evolution in longitudinally expanding
  system~\cite {Kurkela:2018oqw}.}\BibitemShut {Stop}%
\bibitem [{\citenamefont {Kouno}\ \emph {et~al.}(1990)\citenamefont {Kouno},
  \citenamefont {Maruyama}, \citenamefont {Takagi},\ and\ \citenamefont
  {Saito}}]{Kouno:1989ps}%
  \BibitemOpen
  \bibfield  {author} {\bibinfo {author} {\bibfnamefont {H.}~\bibnamefont
  {Kouno}}, \bibinfo {author} {\bibfnamefont {M.}~\bibnamefont {Maruyama}},
  \bibinfo {author} {\bibfnamefont {F.}~\bibnamefont {Takagi}}, \ and\ \bibinfo
  {author} {\bibfnamefont {K.}~\bibnamefont {Saito}},\ }\href {\doibase
  10.1103/PhysRevD.41.2903} {\bibfield  {journal} {\bibinfo  {journal} {Phys.
  Rev.}\ }\textbf {\bibinfo {volume} {D41}},\ \bibinfo {pages} {2903} (\bibinfo
  {year} {1990})}\BibitemShut {NoStop}%
\bibitem [{\citenamefont {Muronga}(2002)}]{Muronga:2001zk}%
  \BibitemOpen
  \bibfield  {author} {\bibinfo {author} {\bibfnamefont {A.}~\bibnamefont
  {Muronga}},\ }\href {\doibase 10.1103/PhysRevLett.89.159901,
  10.1103/PhysRevLett.88.062302} {\bibfield  {journal} {\bibinfo  {journal}
  {Phys. Rev. Lett.}\ }\textbf {\bibinfo {volume} {88}},\ \bibinfo {pages}
  {062302} (\bibinfo {year} {2002})},\ \bibinfo {note} {[Erratum: Phys. Rev.
  Lett.89,159901(2002)]},\ \Eprint {http://arxiv.org/abs/nucl-th/0104064}
  {arXiv:nucl-th/0104064 [nucl-th]} \BibitemShut {NoStop}%
\bibitem [{Note3()}]{Note3}%
  \BibitemOpen
  \bibinfo {note} {Note that the definition of $\tau _\protect \text {hydro}$
  in Eq.~(\ref {eq:hydro}) differs from the one used in Ref.~\cite
  {Kurkela:2018vqr,Kurkela:2018wud}.}\BibitemShut {Stop}%
\bibitem [{\citenamefont {Shen}\ \emph {et~al.}(2016)\citenamefont {Shen},
  \citenamefont {Qiu}, \citenamefont {Song}, \citenamefont {Bernhard},
  \citenamefont {Bass},\ and\ \citenamefont {Heinz}}]{Shen:2014vra}%
  \BibitemOpen
  \bibfield  {author} {\bibinfo {author} {\bibfnamefont {C.}~\bibnamefont
  {Shen}}, \bibinfo {author} {\bibfnamefont {Z.}~\bibnamefont {Qiu}}, \bibinfo
  {author} {\bibfnamefont {H.}~\bibnamefont {Song}}, \bibinfo {author}
  {\bibfnamefont {J.}~\bibnamefont {Bernhard}}, \bibinfo {author}
  {\bibfnamefont {S.}~\bibnamefont {Bass}}, \ and\ \bibinfo {author}
  {\bibfnamefont {U.}~\bibnamefont {Heinz}},\ }\href {\doibase
  10.1016/j.cpc.2015.08.039} {\bibfield  {journal} {\bibinfo  {journal}
  {Comput. Phys. Commun.}\ }\textbf {\bibinfo {volume} {199}},\ \bibinfo
  {pages} {61} (\bibinfo {year} {2016})},\ \Eprint
  {http://arxiv.org/abs/1409.8164} {arXiv:1409.8164 [nucl-th]} \BibitemShut
  {NoStop}%
\bibitem [{\citenamefont {Niemi}\ \emph {et~al.}(2016)\citenamefont {Niemi},
  \citenamefont {Eskola},\ and\ \citenamefont {Paatelainen}}]{Niemi:2015qia}%
  \BibitemOpen
  \bibfield  {author} {\bibinfo {author} {\bibfnamefont {H.}~\bibnamefont
  {Niemi}}, \bibinfo {author} {\bibfnamefont {K.~J.}\ \bibnamefont {Eskola}}, \
  and\ \bibinfo {author} {\bibfnamefont {R.}~\bibnamefont {Paatelainen}},\
  }\href {\doibase 10.1103/PhysRevC.93.024907} {\bibfield  {journal} {\bibinfo
  {journal} {Phys. Rev.}\ }\textbf {\bibinfo {volume} {C93}},\ \bibinfo {pages}
  {024907} (\bibinfo {year} {2016})},\ \Eprint
  {http://arxiv.org/abs/1505.02677} {arXiv:1505.02677 [hep-ph]} \BibitemShut
  {NoStop}%
\bibitem [{\citenamefont {Bazavov}\ \emph {et~al.}(2014)\citenamefont {Bazavov}
  \emph {et~al.}}]{Bazavov:2014pvz}%
  \BibitemOpen
  \bibfield  {author} {\bibinfo {author} {\bibfnamefont {A.}~\bibnamefont
  {Bazavov}} \emph {et~al.} (\bibinfo {collaboration} {HotQCD}),\ }\href
  {\doibase 10.1103/PhysRevD.90.094503} {\bibfield  {journal} {\bibinfo
  {journal} {Phys. Rev.}\ }\textbf {\bibinfo {volume} {D90}},\ \bibinfo {pages}
  {094503} (\bibinfo {year} {2014})},\ \Eprint {http://arxiv.org/abs/1407.6387}
  {arXiv:1407.6387 [hep-lat]} \BibitemShut {NoStop}%
\bibitem [{\citenamefont {Borsanyi}\ \emph {et~al.}(2016)\citenamefont
  {Borsanyi} \emph {et~al.}}]{Borsanyi:2016ksw}%
  \BibitemOpen
  \bibfield  {author} {\bibinfo {author} {\bibfnamefont {S.}~\bibnamefont
  {Borsanyi}} \emph {et~al.},\ }\href {\doibase 10.1038/nature20115} {\bibfield
   {journal} {\bibinfo  {journal} {Nature}\ }\textbf {\bibinfo {volume}
  {539}},\ \bibinfo {pages} {69} (\bibinfo {year} {2016})},\ \Eprint
  {http://arxiv.org/abs/1606.07494} {arXiv:1606.07494 [hep-lat]} \BibitemShut
  {NoStop}%
\bibitem [{\citenamefont {Bernhard}\ \emph {et~al.}(2016)\citenamefont
  {Bernhard}, \citenamefont {Moreland}, \citenamefont {Bass}, \citenamefont
  {Liu},\ and\ \citenamefont {Heinz}}]{Bernhard:2016tnd}%
  \BibitemOpen
  \bibfield  {author} {\bibinfo {author} {\bibfnamefont {J.~E.}\ \bibnamefont
  {Bernhard}}, \bibinfo {author} {\bibfnamefont {J.~S.}\ \bibnamefont
  {Moreland}}, \bibinfo {author} {\bibfnamefont {S.~A.}\ \bibnamefont {Bass}},
  \bibinfo {author} {\bibfnamefont {J.}~\bibnamefont {Liu}}, \ and\ \bibinfo
  {author} {\bibfnamefont {U.}~\bibnamefont {Heinz}},\ }\href {\doibase
  10.1103/PhysRevC.94.024907} {\bibfield  {journal} {\bibinfo  {journal} {Phys.
  Rev.}\ }\textbf {\bibinfo {volume} {C94}},\ \bibinfo {pages} {024907}
  (\bibinfo {year} {2016})},\ \Eprint {http://arxiv.org/abs/1605.03954}
  {arXiv:1605.03954 [nucl-th]} \BibitemShut {NoStop}%
\bibitem [{Note4()}]{Note4}%
  \BibitemOpen
  \bibinfo {note} {For more accurate description, extensions of our study
  including transverse expansion are needed~\cite
  {Kurkela:2018qeb}.}\BibitemShut {Stop}%
\bibitem [{\citenamefont {Muller}\ and\ \citenamefont
  {Rajagopal}(2005)}]{Muller:2005en}%
  \BibitemOpen
  \bibfield  {author} {\bibinfo {author} {\bibfnamefont {B.}~\bibnamefont
  {Muller}}\ and\ \bibinfo {author} {\bibfnamefont {K.}~\bibnamefont
  {Rajagopal}},\ }\bibfield  {booktitle} {\emph {\bibinfo {booktitle}
  {{Proceedings, 1st International Conference on Hard and Electromagnetic
  Probes of High-Energy Nuclear Collisions (Hard Probes 2004): Ericeira,
  Portugal, November 4-10, 2004}}},\ }\href {\doibase
  10.1140/epjc/s2005-02256-3} {\bibfield  {journal} {\bibinfo  {journal} {Eur.
  Phys. J.}\ }\textbf {\bibinfo {volume} {C43}},\ \bibinfo {pages} {15}
  (\bibinfo {year} {2005})},\ \Eprint {http://arxiv.org/abs/hep-ph/0502174}
  {arXiv:hep-ph/0502174 [hep-ph]} \BibitemShut {NoStop}%
\bibitem [{\citenamefont {Hanus}(2018)}]{Hanusthesis}%
  \BibitemOpen
  \bibfield  {author} {\bibinfo {author} {\bibfnamefont {P.}~\bibnamefont
  {Hanus}},\ }\emph {\bibinfo {title} {Entropy in Pb-Pb Collisions at the
  LHC}},\ \href
  {http://www.physi.uni-heidelberg.de//Publications/Bachelor_Thesis_Patrick_Hanus_2018.pdf}
  {\bibinfo {type} {Bachelor's thesis}},\ \bibinfo  {school} {University of
  Heidelberg} (\bibinfo {year} {2018}),\ \bibinfo {note} {supervisor Prof.
  Klaus Reygers}\BibitemShut {NoStop}%
\bibitem [{Note5()}]{Note5}%
  \BibitemOpen
  \bibinfo {note} {Note that the combination $ \left ({\tau }/{\tau _R}\right
  )^{3} \left ({\tau }/{R}\right )^{-2} $ is independent of time.}\BibitemShut
  {Stop}%
\bibitem [{Note6()}]{Note6}%
  \BibitemOpen
  \bibinfo {note} {For example, our model neglects quark masses, which could
  delay flavour equilibration. Even then the relation of strangeness content in
  the QGP and hadronic phase is not trivial~\cite
  {Lee:1987mj,Sollfrank:1995bn,Letessier:1993hi}}\BibitemShut {NoStop}%
\bibitem [{\citenamefont {Citron}\ \emph {et~al.}(2018)\citenamefont {Citron}
  \emph {et~al.}}]{Citron:2018lsq}%
  \BibitemOpen
  \bibfield  {author} {\bibinfo {author} {\bibfnamefont {Z.}~\bibnamefont
  {Citron}} \emph {et~al.},\ }in\ \href@noop {} {\emph {\bibinfo {booktitle}
  {{HL/HE-LHC Workshop: Workshop on the Physics of HL-LHC, and Perspectives at
  HE-LHC Geneva, Switzerland, June 18-20, 2018}}}}\ (\bibinfo {year} {2018})\
  \Eprint {http://arxiv.org/abs/1812.06772} {arXiv:1812.06772 [hep-ph]}
  \BibitemShut {NoStop}%
\bibitem [{\citenamefont {Mrowczynski}(1988)}]{Mrowczynski:1988dz}%
  \BibitemOpen
  \bibfield  {author} {\bibinfo {author} {\bibfnamefont {S.}~\bibnamefont
  {Mrowczynski}},\ }\bibfield  {booktitle} {\emph {\bibinfo {booktitle} {{Arles
  Multipart.Dyn.1988:0499}}},\ }\href {\doibase 10.1016/0370-2693(88)90124-4,
  10.1016/j.physletb.2007.09.039} {\bibfield  {journal} {\bibinfo  {journal}
  {Phys. Lett.}\ }\textbf {\bibinfo {volume} {B214}},\ \bibinfo {pages} {587}
  (\bibinfo {year} {1988})},\ \bibinfo {note} {[Erratum: Phys.
  Lett.B656,273(2007)]}\BibitemShut {NoStop}%
\bibitem [{\citenamefont {Mrowczynski}(1993)}]{Mrowczynski:1993qm}%
  \BibitemOpen
  \bibfield  {author} {\bibinfo {author} {\bibfnamefont {S.}~\bibnamefont
  {Mrowczynski}},\ }\href {\doibase 10.1016/0370-2693(93)91330-P} {\bibfield
  {journal} {\bibinfo  {journal} {Phys. Lett.}\ }\textbf {\bibinfo {volume}
  {B314}},\ \bibinfo {pages} {118} (\bibinfo {year} {1993})}\BibitemShut
  {NoStop}%
\bibitem [{\citenamefont {Mrowczynski}\ and\ \citenamefont
  {Thoma}(2000)}]{Mrowczynski:2000ed}%
  \BibitemOpen
  \bibfield  {author} {\bibinfo {author} {\bibfnamefont {S.}~\bibnamefont
  {Mrowczynski}}\ and\ \bibinfo {author} {\bibfnamefont {M.~H.}\ \bibnamefont
  {Thoma}},\ }\href {\doibase 10.1103/PhysRevD.62.036011} {\bibfield  {journal}
  {\bibinfo  {journal} {Phys. Rev.}\ }\textbf {\bibinfo {volume} {D62}},\
  \bibinfo {pages} {036011} (\bibinfo {year} {2000})},\ \Eprint
  {http://arxiv.org/abs/hep-ph/0001164} {arXiv:hep-ph/0001164 [hep-ph]}
  \BibitemShut {NoStop}%
\bibitem [{\citenamefont {Kurkela}\ and\ \citenamefont
  {Moore}(2011{\natexlab{a}})}]{Kurkela:2011ub}%
  \BibitemOpen
  \bibfield  {author} {\bibinfo {author} {\bibfnamefont {A.}~\bibnamefont
  {Kurkela}}\ and\ \bibinfo {author} {\bibfnamefont {G.~D.}\ \bibnamefont
  {Moore}},\ }\href {\doibase 10.1007/JHEP11(2011)120} {\bibfield  {journal}
  {\bibinfo  {journal} {JHEP}\ }\textbf {\bibinfo {volume} {11}},\ \bibinfo
  {pages} {120} (\bibinfo {year} {2011}{\natexlab{a}})},\ \Eprint
  {http://arxiv.org/abs/1108.4684} {arXiv:1108.4684 [hep-ph]} \BibitemShut
  {NoStop}%
\bibitem [{\citenamefont {Kurkela}\ and\ \citenamefont
  {Moore}(2011{\natexlab{b}})}]{Kurkela:2011ti}%
  \BibitemOpen
  \bibfield  {author} {\bibinfo {author} {\bibfnamefont {A.}~\bibnamefont
  {Kurkela}}\ and\ \bibinfo {author} {\bibfnamefont {G.~D.}\ \bibnamefont
  {Moore}},\ }\href {\doibase 10.1007/JHEP12(2011)044} {\bibfield  {journal}
  {\bibinfo  {journal} {JHEP}\ }\textbf {\bibinfo {volume} {12}},\ \bibinfo
  {pages} {044} (\bibinfo {year} {2011}{\natexlab{b}})},\ \Eprint
  {http://arxiv.org/abs/1107.5050} {arXiv:1107.5050 [hep-ph]} \BibitemShut
  {NoStop}%
\bibitem [{\citenamefont {Mrowczynski}(2002)}]{Mrowczynski:2001az}%
  \BibitemOpen
  \bibfield  {author} {\bibinfo {author} {\bibfnamefont {S.}~\bibnamefont
  {Mrowczynski}},\ }\href {\doibase 10.1103/PhysRevD.65.117501} {\bibfield
  {journal} {\bibinfo  {journal} {Phys. Rev.}\ }\textbf {\bibinfo {volume}
  {D65}},\ \bibinfo {pages} {117501} (\bibinfo {year} {2002})},\ \Eprint
  {http://arxiv.org/abs/hep-ph/0112100} {arXiv:hep-ph/0112100 [hep-ph]}
  \BibitemShut {NoStop}%
\bibitem [{\citenamefont {Schenke}\ and\ \citenamefont
  {Strickland}(2006)}]{Schenke:2006fz}%
  \BibitemOpen
  \bibfield  {author} {\bibinfo {author} {\bibfnamefont {B.}~\bibnamefont
  {Schenke}}\ and\ \bibinfo {author} {\bibfnamefont {M.}~\bibnamefont
  {Strickland}},\ }\href {\doibase 10.1103/PhysRevD.74.065004} {\bibfield
  {journal} {\bibinfo  {journal} {Phys. Rev.}\ }\textbf {\bibinfo {volume}
  {D74}},\ \bibinfo {pages} {065004} (\bibinfo {year} {2006})},\ \Eprint
  {http://arxiv.org/abs/hep-ph/0606160} {arXiv:hep-ph/0606160 [hep-ph]}
  \BibitemShut {NoStop}%
\bibitem [{\citenamefont {Kurkela}\ \emph {et~al.}(2018)\citenamefont
  {Kurkela}, \citenamefont {Wiedemann},\ and\ \citenamefont
  {Wu}}]{Kurkela:2018qeb}%
  \BibitemOpen
  \bibfield  {author} {\bibinfo {author} {\bibfnamefont {A.}~\bibnamefont
  {Kurkela}}, \bibinfo {author} {\bibfnamefont {U.~A.}\ \bibnamefont
  {Wiedemann}}, \ and\ \bibinfo {author} {\bibfnamefont {B.}~\bibnamefont
  {Wu}},\ }\href@noop {} {\  (\bibinfo {year} {2018})},\ \Eprint
  {http://arxiv.org/abs/1805.04081} {arXiv:1805.04081 [hep-ph]} \BibitemShut
  {NoStop}%
\bibitem [{\citenamefont {Lee}\ \emph {et~al.}(1988)\citenamefont {Lee},
  \citenamefont {Rhoades-Brown},\ and\ \citenamefont {Heinz}}]{Lee:1987mj}%
  \BibitemOpen
  \bibfield  {author} {\bibinfo {author} {\bibfnamefont {K.~S.}\ \bibnamefont
  {Lee}}, \bibinfo {author} {\bibfnamefont {M.~J.}\ \bibnamefont
  {Rhoades-Brown}}, \ and\ \bibinfo {author} {\bibfnamefont {U.~W.}\
  \bibnamefont {Heinz}},\ }\href {\doibase 10.1103/PhysRevC.37.1452} {\bibfield
   {journal} {\bibinfo  {journal} {Phys. Rev.}\ }\textbf {\bibinfo {volume}
  {C37}},\ \bibinfo {pages} {1452} (\bibinfo {year} {1988})}\BibitemShut
  {NoStop}%
\bibitem [{\citenamefont {Sollfrank}\ and\ \citenamefont
  {Heinz}(1995)}]{Sollfrank:1995bn}%
  \BibitemOpen
  \bibfield  {author} {\bibinfo {author} {\bibfnamefont {J.}~\bibnamefont
  {Sollfrank}}\ and\ \bibinfo {author} {\bibfnamefont {U.~W.}\ \bibnamefont
  {Heinz}},\ }\href {\doibase 10.1142/9789812830661_0010} {\ ,\ \bibinfo
  {pages} {555} (\bibinfo {year} {1995})},\ \Eprint
  {http://arxiv.org/abs/nucl-th/9505004} {arXiv:nucl-th/9505004 [nucl-th]}
  \BibitemShut {NoStop}%
\bibitem [{\citenamefont {Letessier}\ \emph {et~al.}(1995)\citenamefont
  {Letessier}, \citenamefont {Tounsi}, \citenamefont {Heinz}, \citenamefont
  {Sollfrank},\ and\ \citenamefont {Rafelski}}]{Letessier:1993hi}%
  \BibitemOpen
  \bibfield  {author} {\bibinfo {author} {\bibfnamefont {J.}~\bibnamefont
  {Letessier}}, \bibinfo {author} {\bibfnamefont {A.}~\bibnamefont {Tounsi}},
  \bibinfo {author} {\bibfnamefont {U.~W.}\ \bibnamefont {Heinz}}, \bibinfo
  {author} {\bibfnamefont {J.}~\bibnamefont {Sollfrank}}, \ and\ \bibinfo
  {author} {\bibfnamefont {J.}~\bibnamefont {Rafelski}},\ }\href {\doibase
  10.1103/PhysRevD.51.3408} {\bibfield  {journal} {\bibinfo  {journal} {Phys.
  Rev.}\ }\textbf {\bibinfo {volume} {D51}},\ \bibinfo {pages} {3408} (\bibinfo
  {year} {1995})},\ \Eprint {http://arxiv.org/abs/hep-ph/9212210}
  {arXiv:hep-ph/9212210 [hep-ph]} \BibitemShut {NoStop}%
\end{thebibliography}%

\end{document}